\documentclass[sigconf]{acmart}
\AtBeginDocument{%
  \providecommand\BibTeX{{%
    Bib\TeX}}}

\usepackage{titlesec}

\titlespacing*{\subsection}{0pt}{*1}{*0.5}
\titlespacing*{\subsubsection}{0pt}{*1}{*0.5}
\usepackage{graphicx}
\usepackage{subfiles}
\usepackage{float}
\usepackage{todonotes}
\usepackage{booktabs}
\usepackage{multirow}
\PassOptionsToPackage{hyphens}{url}\usepackage{hyperref}

\usepackage{dsfont}
\usepackage{amsmath}

\usepackage{caption}
\usepackage{subcaption}
\usepackage{enumitem}

\usepackage{nicematrix}
\usepackage[dvipsnames]{xcolor}

\usepackage{pifont}%

\AtBeginDocument{%
  \providecommand\BibTeX{{%
    \normalfont B\kern-0.5em{\scshape i\kern-0.25em b}\kern-0.8em\TeX}}}

\copyrightyear{2026}
\acmYear{2026}
\setcopyright{cc}
\setcctype{by}
\acmConference[ICTIR '26]{Proceedings of the 2026 International ACM SIGIR Conference on Innovative Concepts and Theories in Information Retrieval (ICTIR)}{July 25, 2026}{Melbourne, VIC, Australia}
\acmBooktitle{Proceedings of the 2026 International ACM SIGIR Conference on Innovative Concepts and Theories in Information Retrieval (ICTIR) (ICTIR '26), July 25, 2026, Melbourne, VIC, Australia}
\acmDOI{10.1145/3805713.3820424}
\acmISBN{979-8-4007-2600-2/2026/07}

\begin{document}

\clubpenalties 3 1001 1002 0
\widowpenalties 3 2001 2002 0

\title{Beyond Relevance: On the Relationship Between Retrieval and RAG Information Coverage}

\author{Saron Samuel}
\affiliation{
    \institution{Johns Hopkins University }
    \city{Baltimore}
    \state{MD}
    \country{USA}
}
\email{ssamue21@jhu.edu}
\author{Alexander Martin}
\affiliation{
    \institution{Johns Hopkins University }
    \city{Baltimore}
    \state{MD}
    \country{USA}
}
\email{amart233@jhu.edu}

\author{Eugene Yang}
\affiliation{
  \institution{Johns Hopkins HLTCOE}
  \city{Baltimore}
  \state{MD}
  \country{USA}
}
\email{eugene.yang@jhu.edu}

\author{Andrew Yates}
\affiliation{
  \institution{Johns Hopkins HLTCOE}
  \city{Baltimore}
  \state{MD}
  \country{USA}
}
\email{andrew.yates@jhu.edu}

\author{Dawn Lawrie}
\affiliation{
  \institution{Johns Hopkins HLTCOE}
  \city{Baltimore}
  \state{MD}
  \country{USA}
}
\email{lawrie@jhu.edu}

\author{Ian Soboroff}
\affiliation{
    \institution{National Institute of Standards and Technology}
    \city{Gaithersburg}
    \state{MD}
    \country{USA}
}
\email{ian.soboroff@nist.gov}

\author{Laura Dietz}
\affiliation{
    \institution{University of New Hampshire}
    \city{Durham}
    \state{NH}
    \country{USA}
}

\email{dietz@cs.unh.edu}

\author{Benjamin Van Durme}
\affiliation{
    \institution{Johns Hopkins University }
    \city{Baltimore}
    \state{MD}
    \country{USA}
}
\email{vandurme@jhu.edu}

\renewcommand{\shortauthors}{Saron Samuel et al.}

\begin{abstract}
Retrieval-augmented generation (RAG) systems combine document retrieval with a generative model to address complex information-seeking tasks like report generation. 
While the relationship between retrieval information coverage and generation effectiveness seems intuitive, it has not been systematically studied. We investigate the relationship between upstream retrieval metrics (coverage-based and relevance-based) and the downstream generated response's information coverage.
Through experiments across two text RAG benchmarks (TREC NeuCLIR 2024 and TREC RAG 2024) and one multimodal benchmark (WikiVideo), we analyze 15 text retrieval stacks and 10 multimodal retrieval stacks across four RAG pipelines and two evaluation frameworks (Auto-ARGUE and MiRAGE). Our findings demonstrate consistent positive correlations between coverage-based retrieval metrics and information coverage in generated responses at both topic and system levels. 
This relationship holds most strongly when retrieval and generation objectives are aligned, and weakens as RAG pipelines become more iterative and complex. 
These findings provide empirical support for using coverage-based retrieval metrics as efficient proxies for RAG information coverage evaluation, reducing the need for costly end-to-end evaluation.

\end{abstract}



\maketitle

\section{Introduction}

As information systems have moved from merely providing source information for human consumption to providing a synthesized text that summarizes the source information, system architecture has changed dramatically. 
When the system outputs a ranked list or a search engine result page (SERP), gathering documents that are most likely to be relevant to the user query has been the typical design goal of the systems.
The community generally refers to this class of problems as \textit{ad hoc retrieval}. 

Large language models (LLMs) have shifted user expectations. Rather than browsing a ranked list, users increasingly expect a single, coherent piece of text synthesized from some source information. A succinct and well-organized summary is usually preferable to a simple concatenation of relevant documents~\cite{Zhao_2021}. Since the LLM will synthesize retrieved documents into a single response, documents that cover the same information become redundant after the first has been processed ~\cite{asaiself, duh2025hltcoe-liverag, duh2025hltcoe-gen-team-trec, yang2025hltcoetrec, cheng2024xrag}. Instead, the retrieval system should surface documents that collectively cover multiple aspects or facets of the information need without redundancy. Similar and related information are expected to be grouped, rephrased, or merged~\cite{barzilay-etal-1999-information} and multiple facets should be included if they are all relevant to the user query~\cite{10.1145/3121050.3121099}. 
This kind of retrieval problem is generally being referred to as \textit{report generation}~\cite{lawrie2025overviewtrec2024neuclir}.

These systems are usually implemented with
retrieval-augmented generation (RAG), which combines the strengths of document retrieval with LLM generation to address complex information seeking tasks. 
Ad hoc retrieval, in this case, becomes an upstream component in the system by providing source information to the generation model. 
However, given the final objective of report generation, it \textit{should} be sufficient for the upstream retrieval model to provide a set 
of documents that cover all needed aspects in the final generation. 
To date, LLMs are not able to faithfully leverage the full advertised context window, exhibiting information loss~\cite{liu-etal-2024-lost, du-etal-2025-context} and hallucination~\cite{kalai2025languagemodelshallucinate} when exceeding a certain effective limit. 
Minimizing the number of documents the generation model needs to process \textit{should} provide benefits to the final generated responses. 
The problem of retrieving for multiple aspects with penalization on redundant information has been studied in prior retrieval literature on diversity ranking~\cite{Carbonell2018}, in line with information coverage~\cite{10.1007/s10791-011-9178-4}. 

However, the relationship between information coverage of the upstream retrieval and the final generation quality, though logical, has not been systematically studied.  Recent work has shown that LLMs frequently produce responses that are factually accurate but incomplete in coverage, failing to address the full diversity of aspects relevant to a query ~\cite{salemi2025planandrefinediversecomprehensiveretrievalaugmented, samarinas-etal-2025-beyond}, underscoring the importance of coverage as a first-class evaluation criterion in RAG systems.
Furthermore, 
end-to-end evaluation of generated responses requires running complete RAG pipelines, which incurs substantial computational cost.
Unlike document-level judgments, judgments of generated responses are not easily reusable, so evaluating them incurs substantial further costs in the form of collecting new judgments from a human or LLM.
Moreover, the LLM itself adds variability and noise, as different LLMs and generation strategies can produce divergent outputs even when given identical retrieved context, resulting in noisy signals when attributing effectiveness to each component in the pipeline~\cite{atil2024non, blair2025llms}. 

In this paper, we aim to provide a systematic study on the relationship between upstream retrieval and the downstream generation quality. 
Particularly, we focus on information coverage (usually materialized as nugget coverage in recent literature ~\cite{abbasiantaeb2025conversationalgoldevaluatingpersonalized, pradeep2024initialnuggetevaluationresults}) of the final generation quality since it is the primary purpose of employing a retrieval model in the generation pipeline. 
Other qualities, such as fluency and faithfulness, can already be attributed to the generation stage itself~\cite{mcmillan2025transparentreasoningdrivesfaithfulness}, thus are excluded in this study. 
Particularly, 
we aim to answer the question: 
``What is the relationship between the upstream retrieval metrics (coverage-based and relevance-based) and information coverage of the downstream generation responses in RAG?''
With our empirical evidence, we show that coverage-based retrieval metrics are consistently more predictive of downstream nugget coverage than relevance-based metrics, particularly for complex information needs.
This evidence provides empirical grounds for %
simplifying the evaluation of information coverage to focus on the upstream retrieval model, reducing both computational costs and experimental noise. 

Our analysis operates at multiple levels. 
The topic-level analysis examines whether better ranked list for a specific query leads to a better generated response. 
The system-level analysis assesses whether employing a more effective retrieval model yields a more effective RAG pipeline in general, measured by information coverage. 

Our analysis is guided by five research questions: 
\begin{itemize}
    \item \textbf{RQ1:} For a given RAG pipeline, does a rank list with higher information coverage lead to a better generated response for a specific topic? 
    \item \textbf{RQ2:} Does using a more effective retrieval system lead to a more effective RAG system on average across topics?
    \item \textbf{RQ3:} Can a more complex RAG pipeline compensate for a less effective retrieval system?
    \item \textbf{RQ4:} Do these relationships hold across different automatic RAG evaluators?
    \item \textbf{RQ5: }Do these relationships hold in multimodal RAG? 
\end{itemize}    
We examine RQ1 and RQ2 at both topic and system levels, across in-domain and cross-domain settings, and across simple retrieve-then-generate pipelines ~\cite{yang2025hltcoetrec,dietz2026incorporating,lajewska2025ginger} and complex iterative strategies ~\cite{asaiself, chan2024rqraglearningrefinequeries}. Our contributions are:

\begin{itemize}[leftmargin=*, topsep=5pt]
    \item We demonstrate that nugget-oriented retrieval metrics serve as reliable indicators of RAG information coverage, with consistent positive correlations observed across topic and system levels, and substantially stronger correlations than relevance-based metrics across text and multimodal benchmarks.
    \item We show that the RAG pipeline complexity affects the retrieval-generation relationship. Simpler linear pipelines benefit directly from retrieval improvements, while complex iterative pipelines can partially decouple generation quality from retrieval effectiveness by adapting queries to retrieval system capabilities.
    \item We validate our findings across multiple generation strategies (GPT-Researcher, Bullet List, and LangGraph), evaluation frameworks (Auto-ARGUE and MiRAGE), and modalities (text and video), demonstrating the robustness and generalizability of coverage-based retrieval metrics as proxies for retrieval contributions to RAG performance.
\end{itemize}

\section{Background}

\subsection{Retrieval and Evaluation}
Traditional adhoc retrieval evaluation has focused on document relevance.
Metrics such as MRR, MAP, and nDCG measure ranking quality by how well relevant documents are placed at the top, each assuming a different model of user browsing behavior ~\cite{robertson2008new,10.1145/582415.582418}.

However, retrieval models embedded in a RAG pipeline for complex information seeking tasks
need to retrieve a broad coverage of information to satisfy the complexity of the information need. 
Therefore, coverage metrics, such as Sub-topic Recall~\cite{10.1145/1935826.1935847} and $\alpha$-nDCG~\cite{10.1145/1390334.1390446} become more appropriate, as they evaluate how well a retrieval system gathers information that collectively addresses all aspects of the user's information need ~\cite{10.5555/2070786.2070817}. 
$\alpha$-nDCG~\cite{10.1145/1390334.1390446} incorporates both relevance and diversity, penalizing redundant information while rewarding novel relevant content. Sub-topic recall %
explicitly measures whether retrieval results cover multiple facets of the information needed~\cite{10.1145/1935826.1935847}. These coverage metrics align with the goals of RAG systems, since generation requires comprehensive information to produce complete and accurate responses.

Intent-aware metrics were developed for retrieval diversification tasks, such as TREC Interactive Track~\cite{10.1016/S0306-4573(00)00053-4} and TREC Novelty Track~\cite{soboroff2003overview}. These tasks aim to provide a diverse set of retrieval results covering different potential user intent given a short, ambiguous query. The TREC Complex Answer Retrieval Track~\cite{10.1145/3121050.3121099} requires diversity for answering a simple question with a complex answer, covering multiple facets of the question. 
While metrics from diversification literature are useful, we view diversification as one of the approaches to improve information coverage. 
The NTCIR community offers a parallel family of graded-relevance diversity measures (intent-aware metrics, D-measures) ~\cite{Sakai2021}. While these measures share the same underlying goal of rewarding topical coverage and penalizing redundancy, we use $\alpha$-nDCG as our primary coverage metric because it was originally proposed as a nugget-based measure and maps directly onto the nugget annotation structure of our evaluation datasets, making it the most natural choice for RAG information coverage evaluation.

\subsection{RAG Pipelines} 
\begin{figure}[h]
    \centering
    \includegraphics[width=\linewidth]{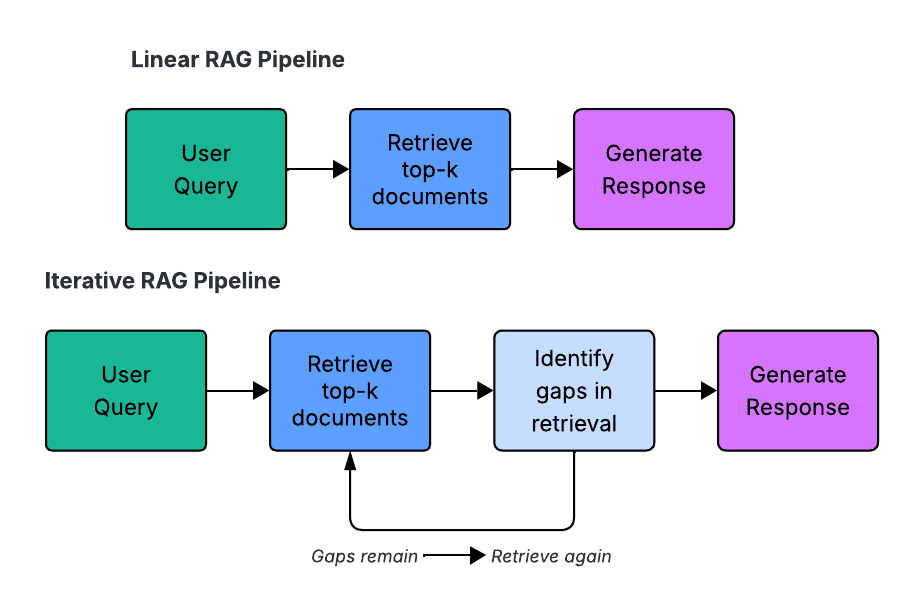}
    \caption{Linear versus iterative RAG pipelines.}
    \label{fig:rag-pipelines}
\end{figure}

RAG systems usually consist of two primary components: an upstream retrieval component that identifies and ranks relevant documents from a corpus and a downstream generation component that synthesizes information from retrieved documents to produce coherent responses. While this basic architecture is consistent across RAG systems, the strategies for combining retrieval and generation vary significantly in complexity ~\cite{gao2024retrievalaugmentedgenerationlargelanguage}, shown in Figure \ref{fig:rag-pipelines}.

\subsubsection{Linear RAG Pipelines. }
The simplest RAG architecture follows a retrieve-then-generate pattern. This is where the system retrieves documents once based on the user's query and then generates a response from the retrieved context. Crucible~\cite{dietz2026incorporating} and GINGER~\cite{lajewska2025ginger} are two examples, where relevant information is extracted from the top-$k$ from a diverse ranking. After retrieval, such systems follow a multi-step response generation stage for the retrieved content via nugget extraction and detection, clustering, ranking, summarization, and fluency enhancement. 

An extension introduces sub-query generation, where the system decomposes complex queries into multiple sub-queries to retrieve more diverse information. GPT Researcher (GPT-R)~\cite{Elovic_gpt-researcher_2023, duh2025hltcoe-liverag} and Bullet List (extractive approach in \citet{yang2025hltcoetrec}) generate multiple queries to gather information from different perspectives before generation. 

\subsubsection{Iterative RAG Pipelines. }
More sophisticated RAG systems employ iterative strategies that alternate between retrieval and generation multiple times, such as Self-RAG, Im-rag, SIM-RAG, etc \cite{asaiself,yang2024rag,yang2025knowing,hayashi2025iterkey,kim2025unirag,fang2025kirag}. 
After an initial retrieval step, the system analyzes retrieved documents to identify information gaps, then performs additional retrieval iterations to fill those gaps. This iterative process continues until the system determines it has sufficient information to generate a comprehensive response.

\subsection{RAG Evaluation}
RAG evaluation presents unique challenges because it assesses generated responses. In the context of tasks like report generation where document citations are required, the evaluation must assess both whether the generated response accurately addresses the user's information needs and whether the generated response reflects and cites information in the retrieved documents. Various evaluation frameworks have emerged to address these challenges, such as ARGUE~\cite{Mayfield_2024}, AutoNuggetizer~\cite{pradeep2024initialnuggetevaluationresults}, EXAM~\cite{farzi2024exambasedevaluationapproachtraditional}, MiRAGE~\cite{martin2025seeingmirageevaluatingmultimodal}, RUBRIC~\cite{10.1145/3664190.3672511}, and ICAT ~\cite{samarinas-etal-2025-beyond}.

Nugget-based evaluation provides a framework for measuring information coverage. Human assessors identify atomic units of information (nuggets) that must be included in a complete answer, then evaluate generated responses based on how many nuggets they contain.

Auto-ARGUE~\cite{walden2025autoargue} applies ARGUE's~\cite{Mayfield_2024} nugget evaluation to RAG systems by representing nuggets as question-answer pairs. Given a query and a set of retrieved documents, assessors create QA pairs that capture distinct pieces of relevant information. Generated responses are then evaluated by checking which QA pairs they answer correctly. This approach provides fine-grained coverage measurement without decomposing the generation into subclaims. Related work on factuality evaluation of long-form LLM generation has independently converged on a similar recall-oriented view, proposing frameworks that measure whether generated responses cover the relevant facts that should be included ~\cite{jafari2026precisionimportanceawarerecallfactuality}, and demonstrating that treating all claims as equally important produces evaluations that are insensitive to errors or omissions in key information ~\cite{wanner2025claimsequalclaimsequal}. Our use of nugget coverage as the primary generation quality metric aligns with this recall-oriented perspective, though we leave importance-weighted coverage as a direction for future work.

MiRAGE~\cite{martin2025seeingmirageevaluatingmultimodal} takes a different approach by representing nuggets as claims that can be decomposed into subclaims. The framework evaluates both whether subclaims are attested in the generated response (recall) and whether attested claims are properly cited with correct sources. MiRAGE's citation recall metric aligns with our goals since a nugget should both appear in the generated response and be grounded in the correct source document.

We use the Auto-ARGUE ~\cite{walden2025autoargue} and MiRAGE ~\cite{martin2025seeingmirageevaluatingmultimodal} to investigate whether retrieval systems that gather more comprehensive information lead to generated responses with higher information coverage. We consider the relationship between information coverage of retrieved documents and of generated responses at both the topic level (i.e., whether improving retrieval coverage on a specific topic leads to better generation coverage on that topic) and at the system level (i.e., whether retrieval systems that have better coverage on average lead to generated responses that have better coverage on average).

\section{Approach}

To systematically investigate the five research questions, 
we conduct our experiments across two text and one multimodal generation task 
to support both within-dataset and cross-dataset analysis.

\subsection{Evaluation Datasets}

For end-to-end RAG evaluation, we use TREC NeuCLIR 2024 Report Generation Pilot Task (NeuCLIR24)~\cite{lawrie2025overviewtrec2024neuclir} and TREC RAG 2024 (RAG24)~\cite{thakur2025supportevaluationtrec2024, pradeep2024initialnuggetevaluationresults} as our two text RAG tasks because of their nugget annotations, which enable information coverage evaluation. 

NeuCLIR24 is a multilingual report generation task where the user input is a problem statement that expresses a rich information need in a paragraph of text, along with a background that describes the profile of the user.
NeuCLIR24 has 19 topics assessed for the report generation pilot. 
The document collection contains more than 10 million news articles extracted from Common Crawl in Chinese, Persian, and Russian. 
RAG24 is a question-answering task that requires the system to retrieve multiple documents as supporting evidence for the final answer. 
RAG24 uses MS MARCO Segment v2.1~\cite{pradeep2024ragnarok} as the document collection with 55 judged queries. 
These questions, for example,\textit{``why are trade-offs so important to the success of a business?''}, while not as simple as factual questions, can often be answered by a single document. However, there are multiple documents that contain acceptable answers that can be retrieved. 

Additionally, we use WikiVideo~\cite{martin2025wikivideoarticlegenerationmultiple}, an event-centric article writing task using video as the supporting documents, to further verify our findings with multimodal RAG. WikiVideo contains 109K videos and 57 topics with an average of 8 relevant videos per topic (maximum 10 videos). 

We refer to the user input for the RAG tasks all as \textit{queries}, and acknowledge the nuanced differences between the kind of user input in each task. 
To evaluate the retrieval models, we use the nuggets in each collection to form a nugget-based qrels that treats each nugget in a topic as an aspect or group.
This format is supported by various retrieval evaluation toolkits, including \texttt{trec-eval}, \texttt{ndeval}, and \texttt{ir-measures}~\cite{macavaney2021streamliningevaluationirmeasures}. 
Additionally, we also include %
MultiVent 2.0 \cite{kriz2025multivent20massivemultilingual}, containing the superset of WikiVideo videos, as additional retrieval tasks for analysis.

\subsection{Retrieval Systems}
We include retrieval systems 
that combine different first-stage retrieval and reranking methods.
For consistency, when reranking, we always rerank the top 100 documents from the first-stage model.
These retrieval stacks are served using RoutIR~\cite{yang2026routirfastservingretrieval}, a serving package that wraps retrieval models with pipeline APIs to support interactions between retrieval and generation systems.  These retrieval systems were selected to represent the major architectural families in retrieval.

For NeuCLIR24 and RAG24, we use the following first-stage retrieval models, covering common first-stage retrieval architectures: 

\begin{itemize}[leftmargin=*]
    \item BM25~\cite{robertson2009probabilistic} as our lexical retrieval baseline. 
    \item PLAID-X \cite{ecir2024translate-distill}, a strong multilingual late-interaction model that achieves the best first-stage effectiveness in TREC NeuCLIR. %
    \item LSR. We use MILCO~\cite{nguyen2025milcolearnedsparseretrieval}, a multilingual learned sparse retrieval (LSR) model that projects text into a shared English lexical space for NeuCLIR24 and SPLADEv3~\cite{spladev3}, a strong English LSR model, for RAG24. 
    \item Qwen3-8B Embed~\cite{zhang2025qwen3embeddingadvancingtext, yang2025qwen3technicalreport}, a multilingual dense embedding model built on the Qwen3.
    \item 3-way RRF, which combines results from PLAID-X, LSR, and Qwen3-8B Embed with Reciprocal Rank Fusion \cite{10.1145/1571941.1572114}
\end{itemize}

To further improve retrieval effectiveness, we employ the two rerankers to rerank the first-stage retrieval results from each method:
\begin{itemize}[leftmargin=*]
    \item Qwen3-8B Reranker \cite{zhang2025qwen3embeddingadvancingtext}, a multilingual pointwise reranking model built on Qwen3 \cite{yang2025qwen3technicalreport}, with state-of-the-art effectiveness in various %
    retrieval tasks.
    \item Rank1-7B \cite{weller2025rank1testtimecomputereranking}, a pointwise reasoning reranking model that allows test-time scaling based on Qwen2.5. 
\end{itemize}

For WikiVideo, we use the following multimodal retrieval models as the first-stage retrieval model that leverage different modalities:

\begin{itemize}[leftmargin=*]
    \item CLIP \cite{radford2021learningtransferablevisualmodels} is a pretrained text-image model. To retrieve videos, we take 16 frames and use the maximum similarity between the query-frame pairs. 
    \item LanguageBind \cite{zhu2024languagebindextendingvideolanguagepretraining}, a multimodal framework that uses language to embed different modalities %
    in a shared space.
    \item MMMORRF \cite{samuel2025mmmorrfmultimodalmultilingualmodularized}, a video retrieval system that combines visual, audio, and text with modality-aware weighted fusion.
    \item Video-ColBERT \cite{reddy2025videocolbertcontextualizedlateinteraction}, a late-interaction text-to-video model.
    \item OmniEmbed \cite{ma2025tevatron20unifieddocument}, an omnimodal encoder on Qwen2.5 Omni~\cite{xu2025qwen25omnitechnicalreport}.
\end{itemize}

For WikiVideo, we employ 
RankVideo \cite{skow2026rankvideoreasoningrerankingtexttovideo}, a reasoning reranker for pointwise reranking in text-to-video retrieval. 

Overall, we investigate 15 retrieval stacks (5 first-stage models with and without the 2 rerankers) for the NeuCLIR24 and RAG24 tasks; 
for WikiVideo, we use 10 stacks.

\subsection{Experiment Setup}
We use four RAG pipelines for NeuCLIR24 and RAG24 tasks: 
GPT-Research (GPT-R)~\cite{Elovic_gpt-researcher_2023, duh2025hltcoe-liverag} with one and three queries, Bullet List~\cite{yang2025hltcoetrec}, and LangGraph~\cite{duh2025hltcoe-gen-team-trec}.
All generation pipelines are given a processing budget of 50 documents to enable a fair comparison between pipelines. 
While sub-query generation and iterative retrieval could be viewed as part of the retrieval stack, we treat them as part of the generation pipeline in this work. This is because in all cases they are driven by the LLM determining what to query, when to query, and how to incorporate retrieved results into the evolving response. Categorizing them as retrieval would conflate the retrieval model's intrinsic effectiveness with the LLM's ability to formulate queries, obscuring the relationship we aim to study.
All text RAG pipelines use Llama-70B-Instruct as the backbone LLM.

GPT-Researcher (GPT-R) \cite{Elovic_gpt-researcher_2023, duh2025hltcoe-liverag} is used as a cascade system that retrieves a seed set of documents given the query, generates additional sub-queries,
retrieves documents with the original query and sub-queries, and then aggregates information into reports. 
We experiment with one query (i.e., no subquery generation) and three queries (i.e., generating two subqueries) with this pipeline. 

Bullet List~\cite{yang2025hltcoetrec} is an extractive system that first generates 10 Google-like queries, retrieves the top 5 documents per query, extracts key related facts from each document, and finally groups the facts with an LLM. The pipeline is implemented with DSPy \cite{khattab2023dspycompilingdeclarativelanguage}. 

Finally, we implement an iterative system with LangGraph \cite{duh2025hltcoe-gen-team-trec}
that iteratively generates sub-queries, retrieves, and compresses documents. It uses reflection to identify knowledge gaps and trigger additional retrieval loops, then drafts and revises responses. Since this pipeline is computationally expensive, we only experiment with NeuCLIR24. 

For WikiVideo, we use CAG \cite{martin2025wikivideoarticlegenerationmultiple} with a Qwen2.5-VL-72B backbone \cite{bai2025qwen25vltechnicalreport}. CAG takes a query and the top 10 videos from the retrieval stack as the input, first extracts key video information related to the query from each video, and then aggregates them into a single response, much like Bullet List but without subquery generation.

\subsection{Retrieval Evaluation}
To study the relationship between retrieval and RAG, we independently evaluate retrieval models outside of the RAG pipelines to assess the system effectiveness. 
We evaluate on both coverage-based and relevance-based metrics.

To measure information coverage, we derive nugget qrels from the nuggets in each RAG collection. Specifically for RAG24, since the nugget-document alignment (i.e., the mapping of nuggets to documents containing them) was never recorded (nor completely assessed), we use Llama-70B-Instruct to judge whether each relevant document contains each nugget in the topic. We acknowledge the potential for circularity here, since the same model family is used in our RAG pipelines. To mitigate this, we verify the LLM-judge against the TREC RAG24 human relevance assessments: treating a document as nugget-relevant if it contains the nugget, our judge achieves a precision of 69\% and a recall of 90\% across all 55 topics. The precision gap reflects the expected mismatch between nugget-level and document-level relevance definitions, TREC assessors labeled documents relevant without requiring them to contain a specific nugget, rather than LLM bias toward any particular retrieval system. We therefore consider the LLM-judged nugget-document alignment sufficiently reliable for evaluating retrieval effectiveness, while acknowledging this as a limitation of the study.

We report three coverage metrics, $\alpha$-nDCG, nDCG using the nugget-based qrels, and Subtopic Recall (StRecall). We use a rank cutoff at 20 for NeuCLIR24 and RAG24, whereas we use a cutoff of 10 for WikiVideo since there are at most 10 relevant videos for each query by design. 

The key differences between these three metrics are the penalty for retrieving redundant nuggets (i.e., nuggets that were already present in another document ranked higher) and how redundancy and document positions affect the score.
$\alpha$-nDCG discounts the gain by a document's rank and reduces the gain when a nugget has already been covered by an earlier document. nDCG uses nugget-based qrels without any consideration of whether a nugget was already covered, which means that only documents that contain a nugget contribute to the score, and there is no penalty for nugget redundancy.
StRecall is a set measure without any penalty on the ranking that measures the fraction of nuggets covered.

Additionally, we report relevance metrics using nDCG with the same cutoff on all datasets. 
We use the qrels that contain the relevance assessments by TREC assessors for relevance-based evaluation on NeuCLIR24 and RAG24, which means that documents that are labeled relevant do not necessarily contain a nugget.
In the tables, we use \textit{(R)} to indicate the nDCG metric reporting is using the relevance-based qrels instead of nugget-based when label sources are not explicitly mentioned for space reasons.

\subsection{RAG Evaluation}

For NeuCLIR24 and RAG24, we use Auto-ARGUE~\cite{walden2025autoargue} as the primary evaluator for the RAG responses. It takes nuggets in the form of question-answer pairs, which natively come with NeuCLIR24 in the dataset. 
For RAG24, where the nuggets are in the form of claims, we again used Llama-70B-Instruct to separate them into questions and answers. To address concerns about LLM-generated evaluation inputs, a human reviewer verified a sample of the resulting QA nuggets and confirmed they faithfully represent the information in the original claims without introducing systematic bias toward any particular generation style. Although Auto-ARGUE scores coverage with an LLM, this circularity is standard in report-generation evaluation, where human annotation at scale is infeasible. The risk of systematic bias is further mitigated by the fact that the nuggets themselves originate from TREC human assessors, not from the LLM.
We evaluate generated responses using Nugget Coverage, which is the proportion of the grounded nuggets covered in the generated response. A nugget is grounded if it is accompanied by a citation to a document containing the nugget.

For WikiVideo, we use MiRAGE ~\cite{martin2025seeingmirageevaluatingmultimodal}, which is its official evaluator. MiRAGE is a nugget-based evaluator for multimodal RAG, measuring factuality, information coverage, and citation support. 
We report Information Precision (InfoP), evaluating factuality, and Information Recall (InfoR), evaluating coverage. 

Note that the definition of InfoR in MiRAGE and Nugget Coverage in Auto-ARGUE is slightly different. 
InfoR directly assesses whether the nugget is covered in the response, where Nugget Coverage rewards only nuggets in the response with appropriate citation. 
This slight difference introduces some nuanced analysis, which we will discuss in the next section. 

To measure the relationship between metrics, we use Pearson correlation coefficient between metrics. 
Since we would like to understand whether improvement in retrieval effectiveness is an indicator of improvement in the final RAG responses, we use Pearson correlation instead of rank correlation to directly capture the value relationship instead of system ranking, which is subject to the choice of and the differences between the systems.

\section{Results}

\begin{table*}[]
\centering

\caption{Retrieval Evaluation on NeuCLIR24 and RAG24. Columns marked with (R) indicate that the qrels for that metric are the relevance-based qrels instead of the nugget-based ones. All metrics in this table use a rank cutoff of 20. }\label{tab:retrieval-results}    
\begin{tabular}{ll|cccc|cccc}
\toprule
 &  & \multicolumn{4}{c|}{NeuCLIR24 Report Generation Pilot Nuggets} & \multicolumn{4}{c}{RAG24 Nuggets} \\
First Stage & Reranker &  $\alpha$-nDCG & nDCG & StRecall & (R) nDCG & $\alpha$-nDCG & nDCG & StRecall & (R) nDCG \\
\midrule
\multirow[l]{3}{*}{BM25} 
 &    --    & 0.349 & 0.328 & 0.545 & 0.170 & 0.450 & 0.207 & 0.708 & 0.263 \\
 & Qwen3-8B & 0.583 & 0.581 & 0.684 & 0.346 & 0.633 & 0.318 & 0.819 & 0.410 \\
 & Rank1-7B & 0.510 & 0.520 & 0.629 & 0.293 & 0.646 & 0.343 & 0.804 & 0.369 \\
\midrule
\multirow[l]{3}{*}{PLAID-X} 
 &    --    & 0.446 & 0.612 & 0.637 & 0.295 & 0.709 & 0.364 & 0.919 & 0.493 \\
 & Qwen3-8B & 0.634 & 0.788 & 0.784 & 0.416 & 0.718 & 0.385 & 0.903 & 0.606 \\
 & Rank1-7B & 0.586 & 0.762 & 0.763 & 0.411 & 0.730 & 0.413 & 0.904 & 0.542 \\
\midrule
\multirow[l]{3}{*}{LSR} 
 &    --    & 0.658 & 0.766 & 0.853 & 0.351 & 0.667 & 0.317 & 0.874 & 0.473 \\
 & Qwen3-8B & 0.720 & 0.892 & 0.874 & 0.496 & 0.697 & 0.365 & 0.883 & 0.578 \\
 & Rank1-7B & 0.651 & 0.877 & 0.848 & 0.458 & 0.733 & 0.412 & 0.908 & 0.537 \\
\midrule
\multirow[l]{3}{*}{\shortstack[l]{Qwen-8B\\Embed}} 
 &    --    & 0.627 & 0.819 & 0.836 & 0.390 & 0.683 & 0.401 & 0.899 & 0.547 \\
 & Qwen3-8B & 0.691 & 0.860 & 0.839 & 0.468 & 0.705 & 0.394 & 0.903 & 0.605 \\
 & Rank1-7B & 0.613 & 0.820 & 0.786 & 0.387 & 0.742 & 0.454 & 0.935 & 0.548 \\
\midrule
\multirow[l]{3}{*}{3 Way RRF} 
 &    --    & 0.637 & 0.796 & 0.851 & 0.374 & 0.705 & 0.366 & 0.899 & 0.587 \\
 & Qwen3-8B & 0.702 & 0.870 & 0.852 & 0.482 & 0.709 & 0.382 & 0.906 & 0.623 \\
 & Rank1-7B & 0.653 & 0.870 & 0.840 & 0.430 & 0.739 & 0.432 & 0.896 & 0.609 \\
\bottomrule
\end{tabular}

\end{table*}

\begin{table*}

\caption{Nugget Coverage of RAG responses on NeuCLIR24 and RAG24 using AutoARGUE.}\label{tab:rag-results}    

\centering
\begin{tabular}{ll|cccc|ccc}
\toprule
 &  & \multicolumn{4}{c|}{NeuCLIR24 Report Generation Pilot} & \multicolumn{3}{c}{RAG24 RAG Task} \\
First Stage & Reranker & GPT-R (1) & GPT-R (3) & Bullet List & LangGraph & GPT-R (1) & GPT-R (3) & Bullet List \\
\midrule
\multirow[l]{3}{*}{BM25} 
 &    --    & 0.449 & 0.527 & 0.570 & 0.490 & 0.496 & 0.512 & 0.602 \\
 & Qwen3-8B & 0.541 & 0.509 & 0.602 & 0.559 & 0.510 & 0.263 & 0.618 \\
 & Rank1-7B & 0.484 & 0.561 & 0.601 & 0.574 & 0.512 & 0.548 & 0.559 \\
\midrule
\multirow[l]{3}{*}{PLAID-X}
 &    --    & 0.504 & 0.523 & 0.586 & 0.545 & 0.599 & 0.588 & 0.647 \\
 & Qwen3-8B & 0.583 & 0.562 & 0.607 & 0.545 & 0.585 & 0.597 & 0.652 \\
 & Rank1-7B & 0.588 & 0.588 & 0.581 & 0.552 & 0.589 & 0.551 & 0.623 \\
\midrule
\multirow[l]{3}{*}{LSR} 
 &    --    & 0.518 & 0.568 & 0.639 & 0.552 & 0.568 & 0.589 & 0.654 \\
 & Qwen3-8B & 0.568 & 0.547 & 0.599 & 0.569 & 0.519 & 0.571 & 0.664 \\
 & Rank1-7B & 0.579 & 0.599 & 0.584 & 0.534 & 0.587 & 0.586 & 0.635 \\
\midrule
\multirow[l]{3}{*}{\shortstack[l]{Qwen-8B\\Embed}} 
 &    --    & 0.587 & 0.582 & 0.608 & 0.560 & 0.592 & 0.603 & 0.662 \\
 & Qwen3-8B & 0.571 & 0.546 & 0.583 & 0.532 & 0.562 & 0.579 & 0.658 \\
 & Rank1-7B & 0.587 & 0.585 & 0.604 & 0.520 & 0.557 & 0.594 & 0.652 \\
\midrule
\multirow[l]{3}{*}{3 Way RRF} 
 &    --    & 0.600 & 0.575 & 0.606 & 0.547 & 0.608 & 0.611 & 0.662 \\
 & Qwen3-8B & 0.611 & 0.599 & 0.597 & 0.546 & 0.582 & 0.620 & 0.662 \\
 & Rank1-7B & 0.582 & 0.550 & 0.610 & 0.490 & 0.570 & 0.579 & 0.649 \\
\bottomrule
\end{tabular}
    
\end{table*}

Table~\ref{tab:retrieval-results} summarizes retrieval effectiveness across the 15 retrieval stacks on NeuCLIR24 and RAG24, measured by four metrics covering both coverage-based and relevance-based evaluation. The stacks span a wide range of quality, from BM25 as the weakest baseline to the Qwen3-8B and Rank1-7B rerankers as the strongest on NeuCLIR24 and RAG24 respectively, providing a broad basis for studying how retrieval quality relates to downstream generation. Table~\ref{tab:rag-results} presents the nugget coverage of generated responses for each of the four RAG pipelines across all 15 retrieval stacks. Comparing across columns (fixing a retrieval stack and varying the pipeline) reveals that pipeline choice meaningfully affects nugget coverage independent of retrieval quality. For example, GPT-R with three queries achieves nugget coverage of 0.263 with the Qwen3-8B reranker on RAG24 but 0.597 with PLAID-X and the same reranker, despite both retrieval stacks being among the stronger performers in Table~\ref{tab:retrieval-results}. This variability illustrates that end-to-end RAG evaluation is noisy. Differences in nugget coverage cannot always be attributed to retrieval quality alone, motivating our focus on the systematic relationship between the two rather than absolute performance values.

\subsection{Topic-Level Analysis}

\begin{figure}
    \centering
    \includegraphics[width=1\linewidth]{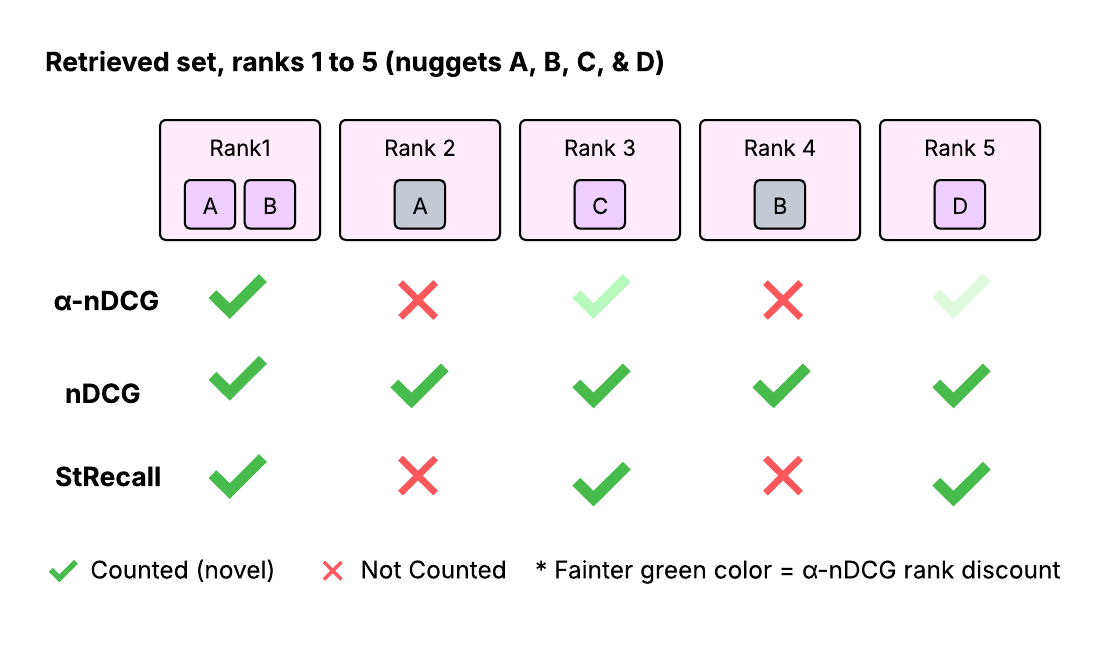}
    \caption{A toy retrieved set of five documents covers nuggets A-D, and ranks 2 and 4 repeat already seen nuggets. $\alpha$-nDCG drops redundant documents and discounts by rank (fainter color). nDCG credits the redundant documents in full. StRecall scores only the union of nuggets, ignoring rank. }
    \label{fig:metrics}
\end{figure}

With all retrieval and generation combinations, we first calculate the topic-level correlation on each benchmark. 
Table~\ref{tab:topic-level-corr} summarizes the Pearson correlation coefficient of the nugget coverage and the retrieval effectiveness using the respective retrieval metric.
Regardless of the generation pipeline, all generated responses correlate with $\alpha$-nDCG using nugget-based labels, the highest (with the exception of Bullet List on NeuCLIR24, where correlation is equally high across all nugget-based metrics), indicating that a RAG pipeline integrating a retrieval system that is capable of retrieving a wide coverage of information is likely to produce a response that also has a high coverage in information. The three coverage metrics differ in how they treat redundancy and rank (Figure~\ref{fig:metrics}), which shapes these correlations.
Nugget-based nDCG@20 also provides a strong indicator. However, since it does not penalize the redundancy of retrieving documents with duplicated information, it may overestimate the usefulness of a document for the downstream generation model in a ranking, thus leading to lower correlation to the nugget coverage. 
Interestingly, StRecall, which is a set-based metric, in general, correlates more strongly with the nugget coverage than nDCG@20. 
While it also does not explicitly penalize the model for including redundant documents, it directly evaluates the set of documents, similar to how the downstream generation model uses the retrieval results. Therefore, this is a metric that aligns with how retrieval results are being used in the pipeline. 
Since each generation pipeline consumes the retrieval results differently, but always takes documents from the top of the ranking, $\alpha$-nDCG provides the most robust evaluation across downstream pipelines.

\begin{table*}[]
\caption{
Topic-level Pearson correlation coefficients between RAG Nugget Coverage and respective retrieval metrics. 
Each value for NeuCLIR24 Pilot and RAG24 is calculated based on 285 (19 topics $\times$ 15 retrieval stacks) and 880 (55 topics $\times$ 15 retrieval stacks) pairs of values, respectively. All metrics in this table use a rank cutoff of 20. 
}
\label{tab:topic-level-corr}
\vspace{-0.5em}
\centering
\begin{tabular}{ll|rrrr|rrr}
\toprule
 & & \multicolumn{4}{c|}{NeuCLIR24 Pilot} & \multicolumn{3}{c}{RAG24} \\
Label & Metric & GPT-R (1) & GPT-R (3) & Bullet List & LangGraph & GPT-R (1) & GPT-R (3) & Bullet List \\
\midrule
Nugget    & $\alpha$-nDCG & 0.5586 & 0.3489 & 0.2645 & 0.3343 & 0.4419 & 0.3785 & 0.3153 \\
Nugget    & nDCG          & 0.4329 & 0.2714 & 0.2623 & 0.1629 & 0.3114 & 0.2564 & 0.1857 \\
Nugget    & StRecall      & 0.4946 & 0.2907 & 0.2694 & 0.2216 & 0.3805 & 0.3231 & 0.2844 \\
\midrule
Relevance & nDCG          & 0.1407 & -0.0131 & 0.0458 & -0.0239 & 0.3467 & 0.3090 & 0.2881 \\
\bottomrule
\end{tabular}
\end{table*}

These three metrics all evaluate the retrieval model on the same information objective, which is coverage. 
We can also compare the correlation of RAG nugget coverage with the relevance-based metric that evaluates the retrieval stacks. 
The last row in Table~\ref{tab:topic-level-corr} presents such correlations. 
Note that although the topics are still paired between retrieval and generation evaluation, the objectives are now different, where the labels used in retrieval evaluation are the relevance of each document.\footnote{NeuCLIR24 explicitly tasked the annotators to assess how useful a particular document is for writing a report without considering each piece of information in the document, while RAG24 simply asked the annotators to assess the relevancy. }
Therefore, this correlation essentially asks the question: Does a retrieval model that retrieves more relevant documents to the topic provide useful information for downstream generation pipelines to improve nugget coverage? 
In NeuCLIR24, where information needs are complex and rich report requests, higher relevance does not indicate higher nugget coverage (low correlations). 
While in RAG24, where information needs are short queries, relevance-based nDCG is still a reasonable indicator for the final nugget coverage, likely because of the kind of information needs are more likely to be satisfied by a single, very relevant document, thus high coverage is not always needed from the retrieval stack.

\begin{table*}[]

\caption{
System-level Pearson correlation between RAG Nugget Coverage and respective retrieval metrics and tasks. 
Cells with a yellow background indicate correlations on the same task (i.e., same objective) and same benchmark; 
ones with a purple background indicate correlations on the same benchmark but different tasks (i.e., different objectives).
All correlations are calculated across RAG pipelines using 15 different retrieval stacks. All metrics in this table use a rank cutoff of 20. 
}\label{tab:system-level-corr}
\vspace{-0.5em}
\centering
\begin{NiceTabular}{ll|cccc|ccc}
\CodeBefore
    \cellcolor{yellow!50}{3-3}\cellcolor{yellow!50}{3-4}\cellcolor{yellow!50}{3-5}\cellcolor{yellow!50}{3-6}
    \cellcolor{yellow!50}{4-3}\cellcolor{yellow!50}{4-4}\cellcolor{yellow!50}{4-5}\cellcolor{yellow!50}{4-6}
    \cellcolor{yellow!50}{5-3}\cellcolor{yellow!50}{5-4}\cellcolor{yellow!50}{5-5}\cellcolor{yellow!50}{5-6}
    \cellcolor{blue!20}{6-3}\cellcolor{blue!20}{6-4}\cellcolor{blue!20}{6-5}\cellcolor{blue!20}{6-6}
    \cellcolor{yellow!50}{7-7}\cellcolor{yellow!50}{7-8}\cellcolor{yellow!50}{7-9}
    \cellcolor{yellow!50}{8-7}\cellcolor{yellow!50}{8-8}\cellcolor{yellow!50}{8-9}
    \cellcolor{yellow!50}{9-7}\cellcolor{yellow!50}{9-8}\cellcolor{yellow!50}{9-9}
    \cellcolor{blue!20}{10-7}\cellcolor{blue!20}{10-8}\cellcolor{blue!20}{10-9}
\Body
\toprule
 &  & \multicolumn{4}{c|}{NeuCLIR24 Pilot} & \multicolumn{3}{c}{RAG24} \\
Retrieval Task & Metric & GPT-R (1) & GPT-R (3) & Bullet List & LangGraph & GPT-R (1) & GPT-R (3) & Bullet List \\
\midrule
\multirow[c]{3}{*}{\shortstack[l]{NeuCLIR24 Pilot\\Report Generation}} 
 & $\alpha$-nDCG & 0.8105 & 0.4894 & 0.4625 & 0.2893 & 0.3543 & 0.2652 & 0.7580 \\
 & nDCG          & 0.8810 & 0.5900 & 0.3346 & 0.1360 & 0.5918 & 0.4855 & 0.9391 \\
 & StRecall      & 0.8251 & 0.5886 & 0.4642 & 0.1608 & 0.5272 & 0.4502 & 0.8459 \\
\midrule
\shortstack{NeuCLIR24 MLIR} 
 & nDCG          & 0.8518 & 0.4960 & 0.1721 & 0.2349 & 0.4061 & 0.3037 & 0.8160 \\
\midrule
\multirow[c]{3}{*}{\shortstack{RAG24 RAG Task}} 
 & $\alpha$-nDCG & 0.7792 & 0.5048 & 0.2963 & 0.2557 & 0.6859 & 0.4027 & 0.8745 \\
 & nDCG          & 0.7953 & 0.5629 & 0.1572 & 0.0726 & 0.6053 & 0.4116 & 0.8391 \\
 & StRecall      & 0.7890 & 0.5161 & 0.2709 & 0.1941 & 0.7923 & 0.5060 & 0.8933 \\
\midrule
RAG24 Retrieval 
 & nDCG          & 0.9028 & 0.4915 & 0.2390 & 0.0941 & 0.6938 & 0.5149 & 0.9130 \\
\bottomrule
\end{NiceTabular}
    
\end{table*}

Overall, to answer our RQ1: \textbf{\textit{For a given RAG pipeline, does an input rank list that covers more information lead to a more effective generated response for a given topic?}} Yes, coverage-based retrieval metrics consistently correlate more strongly with downstream nugget coverage than relevance-based metrics do, particularly for complex information needs such as those in NeuCLIR24 where relevance-based nDCG shows near-zero correlation. For simpler queries like those in RAG24, relevance-based metrics remain a reasonable indicator, but coverage-based metrics still provide a more reliable signal across both datasets. We therefore conclude that input ranked lists with higher information coverage tend to lead to generated responses with higher nugget coverage, and that coverage-based retrieval metrics are better suited than relevance-based ones as proxies for generation quality. We note that topic-level correlations in Table~\ref{tab:topic-level-corr} are moderate in absolute terms (r between 0.26–0.56), reflecting the inherent noise of per-topic RAG evaluation previously noted. The practical value of these correlations lies in their consistency across pipelines and their clear advantage over relevance-based metrics, not in any claim of near-perfect predictability.

\begin{table}[t]
\caption{System-level Pearson correlation between InfoR from MiRAGE and retrieval metrics on NeuCLIR24. The top row additionally reports the macro-averaged InfoR values across all 15 retrieval stacks for each RAG pipeline. 
Retrieval metrics reported in this table use a rank cutoff of 20.}\label{tab:mirage-corr}
\vspace{-0.5em}
\centering
\begin{tabular}{l|cccc}
\toprule
              & GPT-R (1) & GPT-R (3) & Bullet List & LangGraph \\
\midrule
Avg. InfoR    & 0.7279 & 0.7387 & 0.6952 & 0.7325 \\
\midrule
\multicolumn{5}{l}{NeuCLIR24 Pilot}\\
\midrule
$\alpha$-nDCG & 0.6777 & 0.3055 & 0.4623 & 0.2050 \\
nDCG          & 0.6581 & 0.2546 & 0.6290 & 0.1069 \\
StRecall      & 0.5874 & 0.3626 & 0.6050 & 0.2291 \\
\midrule
(R) nDCG      & 0.6869 & 0.2249 & 0.4227 & 0.1434 \\
\midrule
\multicolumn{5}{l}{RAG24}\\
\midrule
$\alpha$-nDCG & 0.6626 & 0.1243 & 0.6121 & -0.0301 \\
nDCG          & 0.7450 & 0.1829 & 0.6047 & -0.2034 \\
StRecall      & 0.5420 & -0.0160 & 0.6484 & -0.0930 \\
\midrule
(R) nDCG      & 0.7100 & 0.1395 & 0.5226 & 0.0364 \\
\bottomrule
\end{tabular}
\end{table}

\begin{table*}[]
\centering

\caption{
Multimodal Retrieval and Report Generation Evaluation Results on WikiVideo using MiRAGE. 
All retrieval metrics use a rank cutoff at 10 since there are at most 10 relevant videos for each topic.}\label{tab:wikivideo-results}    
\vspace{-0.5em}
\begin{tabular}{ll|cc|cccc|cccc}
\toprule
    &  & \multicolumn{2}{c|}{MultiVENT 2.0} & \multicolumn{4}{c|}{WikiVideo Retrieval} &  \multicolumn{2}{c}{WikiVideo Gen} \\
    First Stage & Reranker 
        & Recall & nDCG 
        & $\alpha$-nDCG & nDCG & StRecall & (R) nDCG 
        & InfoP & InfoR \\
\midrule
    \multirow[c]{2}{*}{CLIP} & -- 
        & 0.333 & 0.306
        & 0.538 & 0.498 & 0.724 & 0.498 
        & 81.5 & 89.0 \\
        
        & ReasonRank 
        & 0.477 & 0.478
        & 0.628 & 0.629 & 0.826 & 0.629 
        & 91.9 & 87.6 \\
\midrule
    \multirow[c]{2}{*}{Language Bind} & -- 
        & 0.355 & 0.326
        & 0.476 & 0.457 & 0.634 & 0.457 
        & 84.0 & 89.2 \\
        
        & ReasonRank
        & 0.498 & 0.487
        & 0.566 & 0.563 & 0.754 & 0.563 
        & 94.5 & 86.6 \\
\midrule
    \multirow[c]{2}{*}{Video-ColBERT} & --
        & 0.341 & 0.422
        & 0.431 & 0.383 & 0.634 & 0.383 
        & 83.8 & 89.9 \\
        
        & ReasonRank
        & 0.448 & 0.535
        & 0.553 & 0.522 & 0.734 & 0.522 
        & 87.1 & 87.6 \\
\midrule
    \multirow[c]{2}{*}{OmniEmbed} & --
        & 0.523 & 0.495
        & 0.530 & 0.454 & 0.721 & 0.454 
        & 88.0 & 90.9 \\
        
        & ReasonRank
        & 0.590 & 0.566
        & 0.584 & 0.587 & 0.778 & 0.587 
        & 91.2 & 88.6 \\
\midrule
    \multirow[c]{2}{*}{MMMORRF} & --
        & 0.611 & 0.585
        & 0.540 & 0.503 & 0.724 & 0.503 
        & 94.4 & 88.0 \\
        
        & ReasonRank
        & 0.634 & 0.639
        & 0.605 & 0.617 & 0.785 & 0.617 
        & 91.8 & 88.2 \\
        
\bottomrule
\end{tabular}
\end{table*}

\subsection{System-level Analysis}

While topic-level analysis is useful, it is not practical when designing a RAG pipeline. 
Practitioners often select components based on external benchmark evaluations on queries different from those issued to the actual application. 
Table~\ref{tab:system-level-corr} presents the correlation on the \textit{system} level, where each metric is first averaged over the topics in the respective benchmark before calculating the correlation. 

For the two in-task sets of correlations (yellow blocks), nugget-based retrieval metrics provide higher correlations, indicating that a retrieval system that is generally effective in retrieving wide coverage of information for a specific task would be a good candidate as the upstream retrieval model in a RAG pipeline. 
Compared to topic-level, the relationship shown in Table~\ref{tab:system-level-corr} treats the retrieval metric as a property of the retrieval stack, which is untied from any specific topic. 

We also observe that relevance-based nDCG@20 on RAG24 shows slightly stronger correlation to nugget coverage than $\alpha$-nDCG in some pipeline configurations. This is consistent with the topic-level finding that relevance-based metrics remain informative for shorter, factual queries where a single highly relevant document is often sufficient to cover the information need. However, this advantage does not generalize to NeuCLIR24, where information needs are richer and coverage-based metrics substantially outperform relevance-based ones. This contrast further supports using coverage-based metrics as the more robust choice across task types.

System-level analysis allows us to investigate this relationship across different collections and even different tasks (uncolored blocks in Table~\ref{tab:system-level-corr}). 
When evaluating retrieval on RAG24 RAG task and comparing nugget coverage on NeuCLIR24 (lower left block), it shows lower but still strong correlations compared to evaluating with the same collections. 
If we move even further away to RAG24 Retrieval task (the last row in Table~\ref{tab:system-level-corr}), the correlation becomes lower except for GPT-R with one query. 
This exception is likely due to the nature of GPT-R when using only one query. It ingests the top 50 documents from the single retrieval result, so nDCG at that point essentially evaluates how well the model recalls all the relevant documents, resulting in a very similar correlation as the nugget-based nDCG. 
We observe very similar trends on the RAG24 RAG task (the right portion of Table~\ref{tab:system-level-corr}). 

These results indicate that for the two conditions that we experimented with (retrieval objectives and evaluation), matching both on retrieval and RAG evaluation leads to the strongest correlation. 
Relaxing one gives a similar level of correlation, where the detailed values vary across RAG pipelines. 
When mismatching both, retrieval evaluation becomes less useful as an early indicator of the final RAG nugget coverage, especially for iterative pipelines such as LangGraph, with essentially no correlation (r=0.0941). 
Since the RAG queries in RAG24 are less complex, evaluating retrieval effectiveness from any aspect could provide some level of indication. 

Therefore, to answer RQ2: \textbf{\textit{For a given RAG pipeline, does using a more effective retrieval system as a component lead to a more effective RAG system?}}
Yes, particularly when measuring retrieval systems with a metric that matches the objectives of the final RAG systems. 
Evaluating on different benchmarks or with different objectives %
is still informative, though with lower correlations.

Looking across RAG pipelines, similar to the topic-level analysis, GPT-R with one query shows a strong correlation to the nugget-based retrieval metrics, while using three queries (two generated based on the input queries) starts to deviate. 
Furthermore, Bullet List generates 10 queries from the original user input to try covering various aspects the user may want to see, which exhibit even lower correlation across all retrieval metrics compared to GPT-R on NeuCLIR24. 
In RAG24, again, since the queries are shorter and factual, Bullet List essentially performs query expansion with parametric knowledge in the underlying LLM. 
The net effect becomes query expansion instead of improving information coverage since there are only a few aspects to cover. 
LangGraph has an even more complex interaction with the retrieval model, where it iteratively ingests documents and generates queries to fill in the missing information, leading to lower correlations to the retrieval metrics. 
While LangGraph may not be the most effective RAG pipeline among the pipelines we have included based on Table~\ref{tab:rag-results}, it is the most detached one from the retrieval effectiveness. 

This detachment leads to a different design pattern: the LLM needs to adapt to the retrieval model to issue queries that the retrieval model can understand. 
The focus of development in improving the final generation output moves away from improving the underlying retrieval model, but rather the adaptivity of the LLM to the specific retrieval model. 
This may be preferable for some applications or production environments where the collection is large or only a limited set of retrieval systems is available. 
Based on our experiments, a simpler (or less iterative) RAG pipeline benefits from the improvement of the retrieval model more directly, which may be more preferable for most applications since finetuning LLM for a specific task or use case is more challenging and computationally expensive than adapting a more effective retrieval model, which is usually more economical.

So to answer RQ3: \textbf{\textit{Can a more complex RAG pipeline compensate a less effective retrieval system?}} 
Potentially, but it is not guaranteed. Using a more complex RAG pipeline can detach the final response quality from the effectiveness of the underlying retrieval model, as we have shown with our correlation analysis. However, detachment does not always lead to improvements in the final generation quality, as shown in Table~\ref{tab:rag-results}. 
The key performance bottleneck in this case shifts from retrieving useful documents to better interaction from the LLM to the retrieval model.

\subsection{Using a Different RAG Evaluator}

To further understand the robustness of this relationship, we employ MiRAGE as an alternative RAG evaluator to evaluate the generated responses on the NeuCLIR24 task. 
For brevity, we only present the system-level Pearson correlation between Information Recall (InfoR) from MiRAGE and the retrieval metrics in Table~\ref{tab:mirage-corr}. 
We omit the detailed InfoR scores from the paper and only present the averaged InfoR scores over the 15 retrieval stacks for each RAG pipeline. The overall trend is similar except that MiRAGE prefers LangGraph slightly more than Auto-ARGUE.

Despite slightly different preferences and metric definitions, links between InfoR and nugget-based retrieval metrics stay consistent with the Auto-ARGUE results. When compared with relevance-based nDCG on NeuCLIR, coverage-based metrics keep their edge across RAG pipelines. Notably, when testing retrieval with RAG24 tasks, we still see a high link on GPT-R with one query, but the links on LangGraph drop from slightly positive to zero with AutoARGUE down to zero to slightly negative. This shows that under a more lenient RAG coverage metric (i.e., one that does not ignore data when a sentence lacks citation support), LangGraph output is even more detached from the retrieval model, while other systems still show positive links with retrieval metrics.

Therefore, to answer RQ4: \textbf{\textit{Do these relationships hold across different RAG evaluators?}}
Yes, the relationship between retrieval and RAG information coverage evaluation still holds when using a different evaluator. However, the nuanced difference between evaluators and their metrics may lead to slightly different results since they are measuring the systems from different aspects and based on slightly different criteria.

\subsection{Multimodal RAG}

To further validate our findings, 
in \autoref{tab:wikivideo-results}, we summarize the effectiveness of the multimodal retrieval and generation systems on the WikiVideo video retrieval and article generation task~\cite{martin2025wikivideoarticlegenerationmultiple}. 
The 10 retrieval combinations offer a variety of retrieval quality and variance in both human- and claim-based relevance evaluation. 
Because of the computational cost of multimodal RAG systems, we study only one representative video RAG system using all 10 retrieval stacks.
In Table~\ref{tab:wikivideo-results}, the 10 retrieval stacks exhibit a wide range of retrieval effectiveness on both an external collection (MultiVent 2.0) and the same collection (WikiVideo) as the generation task.

\begin{table}[t]
\caption{System-level Pearson correlation between retrieval metrics and Info metrics on WikiVideo. Retrieval metrics reported in this table use a rank cutoff of 10. }\label{tab:wikivideo-correlation}
    \centering
\vspace{-0.5em}
\begin{tabular}{ll|cc}
\toprule
Retrieval Task & Metric & InfoP & InfoR \\

\midrule
\multirow[c]{4}{*}{\shortstack[l]{WikiVideo\\Retrieval}}
& $\alpha$-nDCG & 0.6476 & -0.5821 \\
& nDCG          & 0.6528 & -0.6825 \\
& StRecall      & 0.6530 & -0.5405 \\
& (R) nDCG      & 0.6528 & -0.6825 \\
\midrule
\multirow[c]{2}{*}{MultiVent 2.0} 
& Recall        & 0.8447 & -0.2837 \\
& nDCG          & 0.7799 & -0.3269 \\
\bottomrule
\end{tabular}
\vspace{-1em}
\end{table}

In \autoref{tab:wikivideo-correlation}, we do a system-level study on the Pearson correlation of MiRAGE's Information Precision (InfoP, for factuality) and Information Recall (InfoR, for coverage) with retrieval quality. Overall, we see a strong link between the factuality of a system and retrieval performance, not information coverage. As shown in \citet{martin2025wikivideoarticlegenerationmultiple}, multimodal language models often overfit on stored knowledge, leading them to rely on it when answering queries rather than using retrieved content, much like LangGraph in the text RAG study. Since WikiVideo topics are major events from 2015 to 2023, data on these events is almost surely part of the pretraining corpus. Retrieval, in this case, serves to verify stored knowledge to boost factuality rather than coverage. This aligns with what we see in Table~\ref{tab:wikivideo-correlation}. Still, retrieval quality remains a strong signal for the quality of generated responses.

Overall, to directly answer our RQ5: \textbf{Do these relationships hold in multimodal RAG?} Yes, with the strong positive correlation between factuality and retrieval effectiveness. 
We expect that when evaluating multimodal RAG benchmarks that actually require gathering information from the collection (which, to our knowledge, does not exist yet), information coverage would again correlate with the retrieval information coverage.

\section{Conclusion}

We show that nugget-based retrieval metrics reliably predict downstream RAG output quality, especially when retrieval and generation objectives align. More complex or iterative pipelines can weaken this link. These findings let developers use retrieval metrics as efficient proxies for generation quality, reducing the need for expensive end-to-end evaluation.

\subsection*{Disclaimer}
Certain products are named in this paper in order to fully specify the experimental procedure adequately. Such mentions should not be taken as endorsement or recommendation of any company, product, or service by NIST, nor are they intended to imply that the products identified are necessarily the best available for this purpose.

\balance
\bibliographystyle{ACM-Reference-Format}
\bibliography{bibio}

@misc{skow2026rankvideoreasoningrerankingtexttovideo,
      title={RANKVIDEO: Reasoning Reranking for Text-to-Video Retrieval}, 
      author={Tyler Skow and Alexander Martin and Benjamin Van Durme and Rama Chellappa and Reno Kriz},
      year={2026},
      eprint={2602.02444},
      archivePrefix={arXiv},
      primaryClass={cs.IR},
      url={https://arxiv.org/abs/2602.02444}, 
}

@inproceedings{
mcmillan2025transparentreasoningdrivesfaithfulness,
title={Towards Transparent Reasoning: What Drives Faithfulness in Large Language Models?},
author={Teague McMillan and Gabriele Dominici and Martin Gjoreski and Marc Langheinrich},
booktitle={NeurIPS 2025 Workshop on Evaluating the Evolving LLM Lifecycle: Benchmarks, Emergent Abilities, and Scaling},
year={2025},
url={https://openreview.net/forum?id=EZ1uEHz4nH}
}

@inproceedings{macavaney2021streamliningevaluationirmeasures,
author = {MacAvaney, Sean and Macdonald, Craig and Ounis, Iadh},
title = {Streamlining Evaluation with ir-measures},
year = {2022},
isbn = {978-3-030-99738-0},
publisher = {Springer-Verlag},
address = {Berlin, Heidelberg},
url = {https://doi.org/10.1007/978-3-030-99739-7_38},
doi = {10.1007/978-3-030-99739-7_38},
booktitle = {Advances in Information Retrieval: 44th European Conference on IR Research, ECIR 2022, Stavanger, Norway, April 10–14, 2022, Proceedings, Part II},
pages = {305–310},
numpages = {6},
location = {Stavanger, Norway}
}

@misc{pradeep2024ragnarok,
      title={Ragnar\"ok: A Reusable RAG Framework and Baselines for TREC 2024 Retrieval-Augmented Generation Track}, 
      author={Ronak Pradeep and Nandan Thakur and Sahel Sharifymoghaddam and Eric Zhang and Ryan Nguyen and Daniel Campos and Nick Craswell and Jimmy Lin},
      year={2024},
      eprint={2406.16828},
      archivePrefix={arXiv},
      primaryClass={cs.IR},
      url={https://arxiv.org/abs/2406.16828}, 
}

@misc{farzi2024exambasedevaluationapproachtraditional,
      title={An Exam-based Evaluation Approach Beyond Traditional Relevance Judgments}, 
      author={Naghmeh Farzi and Laura Dietz},
      year={2024},
      eprint={2402.00309},
      archivePrefix={arXiv},
      primaryClass={cs.IR},
      url={https://arxiv.org/abs/2402.00309}, 
}

@misc{jafari2026precisionimportanceawarerecallfactuality,
      title={Beyond Precision: Importance-Aware Recall for Factuality Evaluation in Long-Form LLM Generation}, 
      author={Nazanin Jafari and James Allan and Mohit Iyyer},
      year={2026},
      eprint={2604.03141},
      archivePrefix={arXiv},
      primaryClass={cs.CL},
      url={https://arxiv.org/abs/2604.03141}, 
}

@misc{wanner2025claimsequalclaimsequal,
      title={All Claims Are Equal, but Some Claims Are More Equal Than Others: Importance-Sensitive Factuality Evaluation of LLM Generations}, 
      author={Miriam Wanner and Leif Azzopardi and Paul Thomas and Soham Dan and Benjamin Van Durme and Nick Craswell},
      year={2025},
      eprint={2510.07083},
      archivePrefix={arXiv},
      primaryClass={cs.CL},
      url={https://arxiv.org/abs/2510.07083}, 
}

@inproceedings{samarinas-etal-2025-beyond,
    title = "Beyond Factual Accuracy: Evaluating Coverage of Diverse Factual Information in Long-form Text Generation",
    author = "Samarinas, Chris  and
      Krubner, Alexander  and
      Salemi, Alireza  and
      Kim, Youngwoo  and
      Zamani, Hamed",
    editor = "Che, Wanxiang  and
      Nabende, Joyce  and
      Shutova, Ekaterina  and
      Pilehvar, Mohammad Taher",
    booktitle = "Findings of the Association for Computational Linguistics: ACL 2025",
    month = jul,
    year = "2025",
    address = "Vienna, Austria",
    publisher = "Association for Computational Linguistics",
    url = "https://aclanthology.org/2025.findings-acl.693/",
    doi = "10.18653/v1/2025.findings-acl.693",
    pages = "13468--13482",
    ISBN = "979-8-89176-256-5",
}

@misc{salemi2025planandrefinediversecomprehensiveretrievalaugmented,
      title={Plan-and-Refine: Diverse and Comprehensive Retrieval-Augmented Generation}, 
      author={Alireza Salemi and Chris Samarinas and Hamed Zamani},
      year={2025},
      eprint={2504.07794},
      archivePrefix={arXiv},
      primaryClass={cs.CL},
      url={https://arxiv.org/abs/2504.07794}, 
}

@inproceedings{10.1145/3664190.3672511,
author = {Farzi, Naghmeh and Dietz, Laura},
title = {Pencils Down! Automatic Rubric-based Evaluation of Retrieve/Generate Systems},
year = {2024},
isbn = {9798400706813},
publisher = {Association for Computing Machinery},
address = {New York, NY, USA},
url = {https://doi.org/10.1145/3664190.3672511},
doi = {10.1145/3664190.3672511},
booktitle = {Proceedings of the 2024 ACM SIGIR International Conference on Theory of Information Retrieval},
pages = {175–184},
numpages = {10},
location = {Washington DC, USA},
series = {ICTIR '24}
}

@misc{gao2024retrievalaugmentedgenerationlargelanguage,
      title={Retrieval-Augmented Generation for Large Language Models: A Survey}, 
      author={Yunfan Gao and Yun Xiong and Xinyu Gao and Kangxiang Jia and Jinliu Pan and Yuxi Bi and Yi Dai and Jiawei Sun and Meng Wang and Haofen Wang},
      year={2024},
      eprint={2312.10997},
      archivePrefix={arXiv},
      primaryClass={cs.CL},
      url={https://arxiv.org/abs/2312.10997}, 
}

@inproceedings{10.1145/3121050.3121099,
author = {Nanni, Federico and Mitra, Bhaskar and Magnusson, Matt and Dietz, Laura},
title = {Benchmark for Complex Answer Retrieval},
year = {2017},
isbn = {9781450344906},
publisher = {Association for Computing Machinery},
address = {New York, NY, USA},
url = {https://doi.org/10.1145/3121050.3121099},
doi = {10.1145/3121050.3121099},
booktitle = {Proceedings of the ACM SIGIR International Conference on Theory of Information Retrieval},
pages = {293–296},
numpages = {4},
location = {Amsterdam, The Netherlands},
series = {ICTIR '17}
}

@inproceedings{soboroff2003overview,
  title={Overview of the TREC 2003 Novelty Track.},
  author={Soboroff, Ian and Harman, Donna and others},
  booktitle={TREC},
  pages={38--53},
  year={2003}
}

@article{10.5555/2070786.2070817,
author = {Chapelle, Olivier and Ji, Shihao and Liao, Ciya and Velipasaoglu, Emre and Lai, Larry and Wu, Su-Lin},
title = {Intent-based diversification of web search results: metrics and algorithms},
year = {2011},
issue_date = {Dec 2011},
publisher = {Kluwer Academic Publishers},
address = {USA},
volume = {14},
number = {6},
issn = {1386-4564},
journal = {Inf. Retr.},
month = dec,
pages = {572–592},
numpages = {21}
}

@inproceedings{10.1145/1935826.1935847,
author = {Clarke, Charles L.A. and Craswell, Nick and Soboroff, Ian and Ashkan, Azin},
title = {A comparative analysis of cascade measures for novelty and diversity},
year = {2011},
isbn = {9781450304931},
publisher = {Association for Computing Machinery},
address = {New York, NY, USA},
url = {https://doi.org/10.1145/1935826.1935847},
doi = {10.1145/1935826.1935847},
booktitle = {Proceedings of the Fourth ACM International Conference on Web Search and Data Mining},
pages = {75–84},
numpages = {10},
location = {Hong Kong, China},
series = {WSDM '11}
}

@inproceedings{10.1145/1390334.1390446,
author = {Clarke, Charles L.A. and Kolla, Maheedhar and Cormack, Gordon V. and Vechtomova, Olga and Ashkan, Azin and B\"{u}ttcher, Stefan and MacKinnon, Ian},
title = {Novelty and diversity in information retrieval evaluation},
year = {2008},
isbn = {9781605581644},
publisher = {Association for Computing Machinery},
address = {New York, NY, USA},
url = {https://doi.org/10.1145/1390334.1390446},
doi = {10.1145/1390334.1390446},
booktitle = {Proceedings of the 31st Annual International ACM SIGIR Conference on Research and Development in Information Retrieval},
pages = {659–666},
numpages = {8},
location = {Singapore, Singapore},
series = {SIGIR '08}
}

@article{10.1145/582415.582418,
author = {J\"{a}rvelin, Kalervo and Kek\"{a}l\"{a}inen, Jaana},
title = {Cumulated gain-based evaluation of IR techniques},
year = {2002},
issue_date = {October 2002},
publisher = {Association for Computing Machinery},
address = {New York, NY, USA},
volume = {20},
number = {4},
issn = {1046-8188},
url = {https://doi.org/10.1145/582415.582418},
doi = {10.1145/582415.582418},
journal = {ACM Trans. Inf. Syst.},
month = oct,
pages = {422–446},
numpages = {25}
}

@conference{chan2024rqraglearningrefinequeries,
title = "RQ-RAG: Learning to Refine Queries for Retrieval Augmented Generation",
author = "Chunpu Xu and Hongyin Luo and Chi-min Chan and Jie Fu and Yike Guo and Wei Xue and Ruibin Yuan",
year = "2024",
language = "English",
note = "Conference on Language Modeling (COLM 2024) ; Conference date: 01-01-2024 Through 01-01-2024",
}

@inproceedings{abbasiantaeb2025conversationalgoldevaluatingpersonalized,
author = {Abbasiantaeb, Zahra and Lupart, Simon and Azzopardi, Leif and Dalton, Jeffrey and Aliannejadi, Mohammad},
title = {Conversational Gold: Evaluating Personalized Conversational Search System Using Gold Nuggets},
year = {2025},
isbn = {9798400715921},
publisher = {Association for Computing Machinery},
address = {New York, NY, USA},
url = {https://doi.org/10.1145/3726302.3730316},
doi = {10.1145/3726302.3730316},
booktitle = {Proceedings of the 48th International ACM SIGIR Conference on Research and Development in Information Retrieval},
pages = {3455–3465},
numpages = {11},
location = {Padua, Italy},
series = {SIGIR '25}
}

@misc{pradeep2024initialnuggetevaluationresults,
      title={Initial Nugget Evaluation Results for the TREC 2024 RAG Track with the AutoNuggetizer Framework}, 
      author={Ronak Pradeep and Nandan Thakur and Shivani Upadhyay and Daniel Campos and Nick Craswell and Jimmy Lin},
      year={2024},
      eprint={2411.09607},
      archivePrefix={arXiv},
      primaryClass={cs.IR},
      url={https://arxiv.org/abs/2411.09607}, 
}

@article{10.1007/s10791-011-9178-4,
author = {Zheng, Wei and Wang, Xuanhui and Fang, Hui and Cheng, Hong},
title = {Coverage-based search result diversification},
year = {2012},
issue_date = {Oct 2012},
publisher = {Kluwer Academic Publishers},
address = {USA},
volume = {15},
number = {5},
issn = {1386-4564},
url = {https://doi.org/10.1007/s10791-011-9178-4},
doi = {10.1007/s10791-011-9178-4},
journal = {Inf. Retr.},
month = oct,
pages = {433–457},
numpages = {25}
}

@article{Carbonell2018,
author = "Jaime G. Carbonell and Jade  Goldstein",
title = "{The Use of MMR and Diversity-Based Reranking in Document Reranking and Summarization}",
year = "2018",
month = "6",
url = "https://kilthub.cmu.edu/articles/journal_contribution/The_Use_of_MMR_and_Diversity-Based_Reranking_in_Document_Reranking_and_Summarization/6610814",
doi = "10.1184/R1/6610814.v1"
}

@misc{kalai2025languagemodelshallucinate,
      title={Why Language Models Hallucinate}, 
      author={Adam Tauman Kalai and Ofir Nachum and Santosh S. Vempala and Edwin Zhang},
      year={2025},
      eprint={2509.04664},
      archivePrefix={arXiv},
      primaryClass={cs.CL},
      url={https://arxiv.org/abs/2509.04664}, 
}

@article{liu-etal-2024-lost,
    title = "Lost in the Middle: How Language Models Use Long Contexts",
    author = "Liu, Nelson F.  and
      Lin, Kevin  and
      Hewitt, John  and
      Paranjape, Ashwin  and
      Bevilacqua, Michele  and
      Petroni, Fabio  and
      Liang, Percy",
    journal = "Transactions of the Association for Computational Linguistics",
    volume = "12",
    year = "2024",
    address = "Cambridge, MA",
    publisher = "MIT Press",
    url = "https://aclanthology.org/2024.tacl-1.9/",
    doi = "10.1162/tacl_a_00638",
    pages = "157--173",
}

@inproceedings{du-etal-2025-context,
    title = "Context Length Alone Hurts {LLM} Performance Despite Perfect Retrieval",
    author = "Du, Yufeng  and
      Tian, Minyang  and
      Ronanki, Srikanth  and
      Rongali, Subendhu  and
      Bodapati, Sravan Babu  and
      Galstyan, Aram  and
      Wells, Azton  and
      Schwartz, Roy  and
      Huerta, Eliu A  and
      Peng, Hao",
    editor = "Christodoulopoulos, Christos  and
      Chakraborty, Tanmoy  and
      Rose, Carolyn  and
      Peng, Violet",
    booktitle = "Findings of the Association for Computational Linguistics: EMNLP 2025",
    month = nov,
    year = "2025",
    address = "Suzhou, China",
    publisher = "Association for Computational Linguistics",
    url = "https://aclanthology.org/2025.findings-emnlp.1264/",
    doi = "10.18653/v1/2025.findings-emnlp.1264",
    pages = "23281--23298",
    ISBN = "979-8-89176-335-7",
}

@inproceedings{barzilay-etal-1999-information,
    title = "Information Fusion in the Context of Multi-Document Summarization",
    author = "Barzilay, Regina  and
      McKeown, Kathleen R.  and
      Elhadad, Michael",
    booktitle = "Proceedings of the 37th Annual Meeting of the Association for Computational Linguistics",
    month = jun,
    year = "1999",
    address = "College Park, Maryland, USA",
    publisher = "Association for Computational Linguistics",
    url = "https://aclanthology.org/P99-1071/",
    doi = "10.3115/1034678.1034760",
    pages = "550--557"
}

@article{Zhao_2021,
   title={QBSUM: A large-scale query-based document summarization dataset from real-world applications},
   volume={66},
   ISSN={0885-2308},
   url={http://dx.doi.org/10.1016/j.csl.2020.101166},
   DOI={10.1016/j.csl.2020.101166},
   journal={Computer Speech \& Language},
   publisher={Elsevier BV},
   author={Zhao, Mingjun and Yan, Shengli and Liu, Bang and Zhong, Xinwang and Hao, Qian and Chen, Haolan and Niu, Di and Long, Bowei and Guo, Weidong},
   year={2021},
   month=mar, pages={101166} }

@misc{bai2025qwen25vltechnicalreport,
      title={Qwen2.5-VL Technical Report}, 
      author={Shuai Bai and Keqin Chen and Xuejing Liu and Jialin Wang and Wenbin Ge and Sibo Song and Kai Dang and Peng Wang and Shijie Wang and Jun Tang and Humen Zhong and Yuanzhi Zhu and Mingkun Yang and Zhaohai Li and Jianqiang Wan and Pengfei Wang and Wei Ding and Zheren Fu and Yiheng Xu and Jiabo Ye and Xi Zhang and Tianbao Xie and Zesen Cheng and Hang Zhang and Zhibo Yang and Haiyang Xu and Junyang Lin},
      year={2025},
      eprint={2502.13923},
      archivePrefix={arXiv},
      primaryClass={cs.CV},
      url={https://arxiv.org/abs/2502.13923}, 
}

@inproceedings{ma2025tevatron20unifieddocument,
author = {Ma, Xueguang and Gao, Luyu and Zhuang, Shengyao and Zhan, Jiaqi Samantha and Callan, Jamie and Lin, Jimmy},
title = {Tevatron 2.0: Unified Document Retrieval Toolkit across Scale, Language, and Modality},
year = {2025},
isbn = {9798400715921},
publisher = {Association for Computing Machinery},
address = {New York, NY, USA},
url = {https://doi.org/10.1145/3726302.3730135},
doi = {10.1145/3726302.3730135},
booktitle = {Proceedings of the 48th International ACM SIGIR Conference on Research and Development in Information Retrieval},
pages = {4061–4065},
numpages = {5},
location = {Padua, Italy},
series = {SIGIR '25}
}

@misc{xu2025qwen25omnitechnicalreport,
      title={Qwen2.5-Omni Technical Report}, 
      author={Jin Xu and Zhifang Guo and Jinzheng He and Hangrui Hu and Ting He and Shuai Bai and Keqin Chen and Jialin Wang and Yang Fan and Kai Dang and Bin Zhang and Xiong Wang and Yunfei Chu and Junyang Lin},
      year={2025},
      eprint={2503.20215},
      archivePrefix={arXiv},
      primaryClass={cs.CL},
      url={https://arxiv.org/abs/2503.20215}, 
}

@inproceedings{radford2021learningtransferablevisualmodels,
  title={Learning Transferable Visual Models From Natural Language Supervision},
  author={Alec Radford and Jong Wook Kim and Chris Hallacy and Aditya Ramesh and Gabriel Goh and Sandhini Agarwal and Girish Sastry and Amanda Askell and Pamela Mishkin and Jack Clark and Gretchen Krueger and Ilya Sutskever},
  booktitle={International Conference on Machine Learning},
  year={2021},
  url={https://api.semanticscholar.org/CorpusID:231591445}
}

@misc{martin2025wikivideoarticlegenerationmultiple,
      title={WikiVideo: Article Generation from Multiple Videos}, 
      author={Alexander Martin and Reno Kriz and William Gantt Walden and Kate Sanders and Hannah Recknor and Eugene Yang and Francis Ferraro and Benjamin Van Durme},
      year={2025},
      eprint={2504.00939},
      archivePrefix={arXiv},
      primaryClass={cs.CV},
      url={https://arxiv.org/abs/2504.00939}, 
}

@INPROCEEDINGS {kriz2025multivent20massivemultilingual,
author = { Kriz, Reno and Sanders, Kate and Etter, David and Murray, Kenton and Carpenter, Cameron and Recknor, Hannah and Guallar-Blasco, Jimena and Martin, Alexander and Yang, Eugene and Van Durme, Benjamin },
booktitle = { 2025 IEEE/CVF Conference on Computer Vision and Pattern Recognition (CVPR) },
title = {{ MultiVENT 2.0: A Massive Multilingual Benchmark for Event-Centric Video Retrieval }},
year = {2025},
volume = {},
ISSN = {},
pages = {24149-24158},
doi = {10.1109/CVPR52734.2025.02249},
url = {https://doi.ieeecomputersociety.org/10.1109/CVPR52734.2025.02249},
publisher = {IEEE Computer Society},
address = {Los Alamitos, CA, USA},
month =Jun}

@inproceedings{Mayfield_2024, series={SIGIR 2024},
   title={On the Evaluation of Machine-Generated Reports},
   url={http://dx.doi.org/10.1145/3626772.3657846},
   DOI={10.1145/3626772.3657846},
   booktitle={Proceedings of the 47th International ACM SIGIR Conference on Research and Development in Information Retrieval},
   publisher={ACM},
   author={Mayfield, James and Yang, Eugene and Lawrie, Dawn and MacAvaney, Sean and McNamee, Paul and Oard, Douglas W. and Soldaini, Luca and Soboroff, Ian and Weller, Orion and Kayi, Efsun and Sanders, Kate and Mason, Marc and Hibbler, Noah},
   year={2024},
   month=jul, pages={1904–1915},
   collection={SIGIR 2024} }

@misc{martin2025seeingmirageevaluatingmultimodal,
      title={Seeing Through the MiRAGE: Evaluating Multimodal Retrieval Augmented Generation}, 
      author={Alexander Martin and William Walden and Reno Kriz and Dengjia Zhang and Kate Sanders and Eugene Yang and Chihsheng Jin and Benjamin Van Durme},
      year={2025},
      eprint={2510.24870},
      archivePrefix={arXiv},
      primaryClass={cs.CL},
      url={https://arxiv.org/abs/2510.24870}, 
}

@inproceedings{yang2026routirfastservingretrieval,
  author       = {Eugene Yang and
                  Andrew Yates and
                  Dawn J. Lawrie and
                  James Mayfield and
                  Trevor Adriaanse},
  editor       = {Ricardo Campos and
                  Adam Jatowt and
                  Yanyan Lan and
                  Mohammad Aliannejadi and
                  Christine Bauer and
                  Sean MacAvaney and
                  Avishek Anand and
                  Zhaochun Ren and
                  Suzan Verberne and
                  Nan Bai and
                  Masoud Mansoury},
  title        = {RoutIR: Fast Serving of Retrieval Pipelines for Retrieval-Augmented
                  Generation},
  booktitle    = {Advances in Information Retrieval - 48th European Conference on Information
                  Retrieval, {ECIR} 2026, Delft, The Netherlands, March 29 - April 2,
                  2026, Proceedings, Part {IV}},
  series       = {Lecture Notes in Computer Science},
  pages        = {578--593},
  publisher    = {Springer},
  year         = {2026},
  url          = {https://doi.org/10.1007/978-3-032-21321-1\_60},
  doi          = {10.1007/978-3-032-21321-1\_60},
  bibsource    = {dblp computer science bibliography, https://dblp.org}
}

@inproceedings{
khattab2023dspycompilingdeclarativelanguage,
title={{DSP}y: Compiling Declarative Language Model Calls into State-of-the-Art Pipelines},
author={Omar Khattab and Arnav Singhvi and Paridhi Maheshwari and Zhiyuan Zhang and Keshav Santhanam and Sri Vardhamanan A and Saiful Haq and Ashutosh Sharma and Thomas T. Joshi and Hanna Moazam and Heather Miller and Matei Zaharia and Christopher Potts},
booktitle={The Twelfth International Conference on Learning Representations},
year={2024},
url={https://openreview.net/forum?id=sY5N0zY5Od}
}

@inproceedings{
zhu2024languagebindextendingvideolanguagepretraining,
title={LanguageBind: Extending Video-Language Pretraining to N-modality by Language-based Semantic Alignment},
author={Bin Zhu and Bin Lin and Munan Ning and Yang Yan and Jiaxi Cui and WANG HongFa and Yatian Pang and Wenhao Jiang and Junwu Zhang and Zongwei Li and Cai Wan Zhang and Zhifeng Li and Wei Liu and Li Yuan},
booktitle={The Twelfth International Conference on Learning Representations},
year={2024},
url={https://openreview.net/forum?id=QmZKc7UZCy}
}

@inproceedings{
weller2025rank1testtimecomputereranking,
title={Rank1: Test-Time Compute for Reranking in Information Retrieval},
author={Orion Weller and Kathryn Ricci and Eugene Yang and Andrew Yates and Dawn Lawrie and Benjamin Van Durme},
booktitle={Second Conference on Language Modeling},
year={2025},
url={https://openreview.net/forum?id=Pg0PAvbhGv}
}

@misc{zhang2025qwen3embeddingadvancingtext,
      title={Qwen3 Embedding: Advancing Text Embedding and Reranking Through Foundation Models}, 
      author={Yanzhao Zhang and Mingxin Li and Dingkun Long and Xin Zhang and Huan Lin and Baosong Yang and Pengjun Xie and An Yang and Dayiheng Liu and Junyang Lin and Fei Huang and Jingren Zhou},
      year={2025},
      eprint={2506.05176},
      archivePrefix={arXiv},
      primaryClass={cs.CL},
      url={https://arxiv.org/abs/2506.05176}, 
}

@InProceedings{reddy2025videocolbertcontextualizedlateinteraction,
    author    = {Reddy, Arun and Martin, Alexander and Yang, Eugene and Yates, Andrew and Sanders, Kate and Murray, Kenton and Kriz, Reno and de Melo, Celso M. and Van Durme, Benjamin and Chellappa, Rama},
    title     = {Video-ColBERT: Contextualized Late Interaction for Text-to-Video Retrieval},
    booktitle = {Proceedings of the IEEE/CVF Conference on Computer Vision and Pattern Recognition (CVPR)},
    month     = {June},
    year      = {2025},
    pages     = {19691-19701}
}

@inproceedings{samuel2025mmmorrfmultimodalmultilingualmodularized,
author = {Samuel, Saron and DeGenaro, Dan and Guallar-Blasco, Jimena and Sanders, Kate and Eisape, Seun and Reddy, Arun and Martin, Alexander and Yates, Andrew and Yang, Eugene and Carpenter, Cameron and Etter, David and Kayi, Efsun and Wiesner, Matthew and Murray, Kenton and Kriz, Reno},
title = {MMMORRF: Multimodal Multilingual MOdularized Reciprocal Rank Fusion},
year = {2025},
isbn = {9798400715921},
publisher = {Association for Computing Machinery},
address = {New York, NY, USA},
url = {https://doi.org/10.1145/3726302.3730157},
doi = {10.1145/3726302.3730157},
booktitle = {Proceedings of the 48th International ACM SIGIR Conference on Research and Development in Information Retrieval},
pages = {4004–4009},
numpages = {6},
location = {Padua, Italy},
series = {SIGIR '25}
}

@inproceedings{10.1145/1571941.1572114,
author = {Cormack, Gordon V. and Clarke, Charles L A and Buettcher, Stefan},
title = {Reciprocal rank fusion outperforms condorcet and individual rank learning methods},
year = {2009},
isbn = {9781605584836},
publisher = {Association for Computing Machinery},
address = {New York, NY, USA},
url = {https://doi.org/10.1145/1571941.1572114},
doi = {10.1145/1571941.1572114},
booktitle = {Proceedings of the 32nd International ACM SIGIR Conference on Research and Development in Information Retrieval},
pages = {758–759},
numpages = {2},
location = {Boston, MA, USA},
series = {SIGIR '09}
}

@misc{yang2025qwen3technicalreport,
      title={Qwen3 Technical Report}, 
      author={An Yang and Anfeng Li and Baosong Yang and Beichen Zhang and Binyuan Hui and Bo Zheng and Bowen Yu and Chang Gao and Chengen Huang and Chenxu Lv and Chujie Zheng and Dayiheng Liu and Fan Zhou and Fei Huang and Feng Hu and Hao Ge and Haoran Wei and Huan Lin and Jialong Tang and Jian Yang and Jianhong Tu and Jianwei Zhang and Jianxin Yang and Jiaxi Yang and Jing Zhou and Jingren Zhou and Junyang Lin and Kai Dang and Keqin Bao and Kexin Yang and Le Yu and Lianghao Deng and Mei Li and Mingfeng Xue and Mingze Li and Pei Zhang and Peng Wang and Qin Zhu and Rui Men and Ruize Gao and Shixuan Liu and Shuang Luo and Tianhao Li and Tianyi Tang and Wenbiao Yin and Xingzhang Ren and Xinyu Wang and Xinyu Zhang and Xuancheng Ren and Yang Fan and Yang Su and Yichang Zhang and Yinger Zhang and Yu Wan and Yuqiong Liu and Zekun Wang and Zeyu Cui and Zhenru Zhang and Zhipeng Zhou and Zihan Qiu},
      year={2025},
      eprint={2505.09388},
      archivePrefix={arXiv},
      primaryClass={cs.CL},
      url={https://arxiv.org/abs/2505.09388}, 
}

@misc{nguyen2025milcolearnedsparseretrieval,
      title={Milco: Learned Sparse Retrieval Across Languages via a Multilingual Connector}, 
      author={Thong Nguyen and Yibin Lei and Jia-Huei Ju and Eugene Yang and Andrew Yates},
      year={2025},
      eprint={2510.00671},
      archivePrefix={arXiv},
      primaryClass={cs.IR},
      url={https://arxiv.org/abs/2510.00671}, 
}

@inproceedings{ecir2024translate-distill,
  author = {Eugene Yang and Dawn Lawrie and James Mayfield and Douglas W. Oard and Scott Miller},
  title = {Translate-Distill: Learning Cross-Language Dense Retrieval by Translation and Distillation},
  booktitle = {Proceedings of the 46th European Conference on Information Retrieval (ECIR)},
  year = {2024},
  url = {https://arxiv.org/abs/2401.04810}
}

@misc{lawrie2025overviewtrec2024neuclir,
      title={Overview of the TREC 2024 NeuCLIR Track}, 
      author={Dawn Lawrie and Sean MacAvaney and James Mayfield and Paul McNamee and Douglas W. Oard and Luca Soldaini and Eugene Yang},
      year={2025},
      eprint={2509.14355},
      archivePrefix={arXiv},
      primaryClass={cs.IR},
      url={https://arxiv.org/abs/2509.14355}, 
}

@inproceedings{dietz2026incorporating,
  title={Incorporating {Q}\&{A} Nuggets into Retrieval-Augmented Generation},
  author={Dietz, Laura and Li, Bryan and Liu, Gabrielle and Ju, Jia-Huei and Yang, Eugene and Lawrie, Dawn and Walden, William and Mayfield, James},
  booktitle={Proceedings of the 48th European Conference on Information Retrieval (ECIR 2026)},
  year={2026}
}

@inproceedings{duh2025hltcoe-gen-team-trec,
  title={{HLTCOE} Generation Team at TREC 2025},
  author={Kevin Duh and Dawn Lawrie and Debashish Chakraborty and Roxana Petcu and Eugene Yang and Kenton Murraya and Daniel Khashabi and Maxime Dassen},
  booktitle={The Thirty-Fourth Text REtrieval Conference Proceedings (TREC2025)},
  year={2025},
  url={https://trec-ragtime.github.io/assets/notebooks/2025/hltcoe-gen.pdf}
}

@misc{yang2025hltcoetrec,
      title={{HLTCOE} at {TREC} 2024 {NeuCLIR} Track}, 
      author={Eugene Yang and Dawn Lawrie and Orion Weller and James Mayfield},
      year={2025},
      eprint={2510.00143},
      archivePrefix={arXiv},
      primaryClass={cs.CL},
      url={https://arxiv.org/abs/2510.00143}, 
}

@article{walden2025autoargue,
  title={{Auto-ARGUE}: {LLM}-Based Report Generation Evaluation},
  author={William Walden and Orion Weller and Laura Dietz and Bryan Li and Gabrielle Kaili-May Liu and Yu Hou and Eugene Yang
},
  journal={arXiv preprint arXiv:2509.26184},
  year={2025}
}

@article{duh2025hltcoe-liverag,
  title={{HLTCOE} at {LiveRAG}: {GPT-Researcher} using {ColBERT} retrieval},
  author={Duh, Kevin and Yang, Eugene and Weller, Orion and Yates, Andrew and Lawrie, Dawn},
  journal={arXiv preprint arXiv:2506.22356},
  year={2025}
}

@misc{Elovic_gpt-researcher_2023,
author = {Elovic, Assaf},
month = jul,
title = {{gpt-researcher}},
url = {https://github.com/assafelovic/gpt-researcher},
version = {0.5.4},
year = {2023}
}

@inproceedings{Sakai2021,
  author    = {Sakai, Tetsuya and Tao, Sijie and Zeng, Zhaohao},
  booktitle = {Proceedings of the 44th {International} {ACM} {SIGIR} {Conference} on {Research} and {Development} in {Information} {Retrieval}},
  pages     = {2376--2382},
  title     = {{WWW3E8}: 259,000 {Relevance} {Labels} for {Studying} the {Effect} of {Document} {Presentation} {Order} for {Relevance} {Assessors}},
  year      = {2021}
}

@inproceedings{thakur2025supportevaluationtrec2024,
author = {Thakur, Nandan and Pradeep, Ronak and Upadhyay, Shivani and Campos, Daniel and Craswell, Nick and Soboroff, Ian and Dang, Hoa Trang and Lin, Jimmy},
title = {Assessing Support for the TREC 2024 RAG Track: A Large-Scale Comparative Study of LLM and Human Evaluations},
year = {2025},
isbn = {9798400715921},
publisher = {Association for Computing Machinery},
address = {New York, NY, USA},
url = {https://doi.org/10.1145/3726302.3730165},
doi = {10.1145/3726302.3730165},
booktitle = {Proceedings of the 48th International ACM SIGIR Conference on Research and Development in Information Retrieval},
pages = {2759–2763},
numpages = {5},
location = {Padua, Italy},
series = {SIGIR '25}
}

@inproceedings{lajewska2025ginger,
  author    = {Weronika Lajewska and Krisztian Balog},
  title     = {GINGER: Grounded Information Nugget-Based Generation of Responses},
  booktitle = {Proceedings of the 48th International ACM SIGIR Conference (SIGIR ’25)},
  year      = {2025},
  note      = {SIGIR 2025 paper},
  url       = {https://krisztianbalog.com/files/sigir2025-ginger.pdf}
}

@inproceedings{asaiself,
  title     = {Self-RAG: Learning to Retrieve, Generate, and Critique through Self-Reflection},
  author    = {Asai, Akari and Wu, Zeqiu and Wang, Yizhong and Sil, Avirup and Hajishirzi, Hannaneh},
  booktitle = {Proceedings of the International Conference on Learning Representations (ICLR)},
  year      = {2024},
  url       = {https://proceedings.iclr.cc/paper_files/paper/2024/file/25f7be9694d7b32d5cc670927b8091e1-Paper-Conference.pdf}
}

@inproceedings{yang2024rag,
  title={Im-rag: Multi-round retrieval-augmented generation through learning inner monologues},
  author={Yang, Diji and Rao, Jinmeng and Chen, Kezhen and Guo, Xiaoyuan and Zhang, Yawen and Yang, Jie and Zhang, Yi},
  booktitle={Proceedings of the 47th International ACM SIGIR Conference on Research and Development in Information Retrieval},
  pages={730--740},
  year={2024}
}

@inproceedings{cheng2024xrag,
  title        = {xRAG: Extreme Context Compression for Retrieval-Augmented Generation with One Token},
  author       = {Cheng, Xin and Wang, Xun and Zhang, Xingxing and Ge, Tao and Chen, Si-Qing and Wei, Furu and Zhang, Huishuai and Zhao, Dongyan},
  booktitle    = {Advances in Neural Information Processing Systems (NeurIPS)},
  year         = {2024}
}

@inproceedings{robertson2008new,
  title={A new interpretation of average precision},
  author={Robertson, Stephen},
  booktitle={Proceedings of the 31st annual international ACM SIGIR conference on Research and development in information retrieval},
  pages={689--690},
  year={2008}
}

@inproceedings{yang2025knowing,
  title={Knowing You Don't Know: Learning When to Continue Search in Multi-round RAG through Self-Practicing},
  author={Yang, Diji and Zeng, Linda and Rao, Jinmeng and Zhang, Yi},
  booktitle={Proceedings of the 48th International ACM SIGIR Conference on Research and Development in Information Retrieval},
  pages={1305--1315},
  year={2025}
}

@inproceedings{kim2025unirag,
  title={UniRAG: A Unified RAG Framework for Knowledge-Intensive Queries with Decomposition, Break-Down Reasoning, and Iterative Rewriting},
  author={Kim, Gun Il and Kim, Jong Wook and Jang, Beakcheol},
  booktitle={Findings of the Association for Computational Linguistics: EMNLP 2025},
  pages={18795--18810},
  year={2025}
}

@inproceedings{fang2025kirag,
    title = "{K}i{RAG}: Knowledge-Driven Iterative Retriever for Enhancing Retrieval-Augmented Generation",
    author = "Fang, Jinyuan  and
      Meng, Zaiqiao  and
      MacDonald, Craig",
    editor = "Che, Wanxiang  and
      Nabende, Joyce  and
      Shutova, Ekaterina  and
      Pilehvar, Mohammad Taher",
    booktitle = "Proceedings of the 63rd Annual Meeting of the Association for Computational Linguistics (Volume 1: Long Papers)",
    month = jul,
    year = "2025",
    address = "Vienna, Austria",
    publisher = "Association for Computational Linguistics",
    url = "https://aclanthology.org/2025.acl-long.929/",
    doi = "10.18653/v1/2025.acl-long.929",
    pages = "18969--18985",
    ISBN = "979-8-89176-251-0",
}

@inproceedings{hayashi2025iterkey,
  title={Iterkey: Iterative keyword generation with llms for enhanced retrieval augmented generation},
  author={Hayashi, Kazuki and Kamigaito, Hidetaka and Kouda, Shinya and Watanabe, Taro},
  booktitle = {Proceedings of the Second Conference on Language Modeling},
  series = {COLM'25},
  year={2025}
}

@article{spladev3,
  title={SPLADE-v3: New baselines for SPLADE},
  author={Lassance, Carlos and D{\'e}jean, Herv{\'e} and Formal, Thibault and Clinchant, St{\'e}phane},
  journal={arXiv preprint arXiv:2403.06789},
  year={2024}
}

@article{robertson2009probabilistic,
  title={The probabilistic relevance framework: BM25 and beyond},
  author={Robertson, Stephen and Zaragoza, Hugo and others},
  journal={Foundations and trends{\textregistered} in information retrieval},
  volume={3},
  number={4},
  pages={333--389},
  year={2009},
  publisher={Now Publishers, Inc.}
}

@inproceedings{atil2024non,
    title = "Non-Determinism of ``Deterministic'' {LLM} System Settings in Hosted Environments",
    author = "At{\i}l, Berk  and
      Aykent, Sarp  and
      Chittams, Alexa  and
      Fu, Lisheng  and
      Passonneau, Rebecca J.  and
      Radcliffe, Evan  and
      Rajagopal, Guru Rajan  and
      Sloan, Adam  and
      Tudrej, Tomasz  and
      Ture, Ferhan  and
      Wu, Zhe  and
      Xu, Lixinyu  and
      Baldwin, Breck",
    editor = "Akter, Mousumi  and
      Chowdhury, Tahiya  and
      Eger, Steffen  and
      Leiter, Christoph  and
      Opitz, Juri  and
      {\c{C}}ano, Erion",
    booktitle = "Proceedings of the 5th Workshop on Evaluation and Comparison of NLP Systems",
    month = dec,
    year = "2025",
    address = "Mumbai, India",
    publisher = "Association for Computational Linguistics",
    url = "https://aclanthology.org/2025.eval4nlp-1.12/",
    doi = "10.18653/v1/2025.eval4nlp-1.12",
    pages = "135--148",
    ISBN = "979-8-89176-305-0",
}

@inproceedings{blair2025llms,
  title={Llms provide unstable answers to legal questions},
  author={Blair-Stanek, Andrew and Van Durme, Benjamin},
  booktitle={Proceedings of the Twentieth International Conference on Artificial Intelligence and Law},
  pages={425--429},
  year={2025}
}

\end{document}